\documentclass[unsortedaddress,superscriptaddress,prl,twocolumn]{revtex4-2}
\usepackage{hyperref}
\usepackage{epsfig}
\usepackage{graphicx}
\usepackage{subfigure}
\usepackage{latexsym}
\usepackage{color}
\usepackage{fullpage}
\usepackage{dcolumn}
\usepackage{bm}
\usepackage[normalem]{ulem}
\usepackage{units}
\usepackage{amsmath}
\usepackage{ textcomp }
\usepackage[paperwidth=210mm,paperheight=297mm,centering,hmargin=2cm,vmargin=2.3cm]{geometry}

\begin{document}

%\title{Impact of the bending energy on the moir\'{e} lattices of two-dimensional materials on substrates}

\title{Pressure evolution of electronic and crystal structure of non-centrosymmetric EuCoGe$_3$}

\author{N. S. Dhami}
\email{nsdhami@ifs.hr}
\affiliation{Institute of Physics, Bijeni\v{c}ka cesta 46, 10000, Zagreb, Croatia}

\author{V. Balédent}
\affiliation{Université Paris-Saclay, CNRS, Laboratoire de Physique des Solides, 91405 Orsay, France}

\author{O. Bednarchuk}
\affiliation{Institute of Low Temperature and Structure Research, Polish Academy of Sciences, Okólna 2, 50-422 Wrocław, Poland}

\author{D. Kaczorowski}
\affiliation{Institute of Low Temperature and Structure Research, Polish Academy of Sciences, Okólna 2, 50-422 Wrocław, Poland}

\author{S. R. Shieh}
\affiliation{Department of Earth Sciences, Department of Physics and Astronomy, University of Western Ontario, London, Ontario N6A-5B7, Canada}

\author{J. M. Ablett}
\affiliation{Synchrotron SOLEIL, L’Orme desMerisiers BP48, St Aubin, 91192 Gif-sur-Yvette, France}

\author{J.-P. Rueff}
\affiliation{Synchrotron SOLEIL, L’Orme desMerisiers BP48, St Aubin, 91192 Gif-sur-Yvette, France}
\affiliation{Laboratoire de Chimie Physique-Matière et Rayonnement, Sorbonne Université, CNRS, 75005 Paris, France}

\author{J. P. Itié}
\affiliation{Synchrotron SOLEIL, L’Orme desMerisiers BP48, St Aubin, 91192 Gif-sur-Yvette, France}

\author{C. M. N. Kumar}
\affiliation{Institute of Physics, Bijeni\v{c}ka cesta 46, 10000, Zagreb, Croatia}

\author{Y. Utsumi}
\email{yutsumi@ifs.hr}
\affiliation{Institute of Physics, Bijeni\v{c}ka cesta 46, 10000, Zagreb, Croatia}

%\alsoaffiliation{Institute of Solid State Physics, Vienna University of Technology,	Wiedner Hauptstra{\ss}e 8-10, 1040 Vienna, Austria}
%\alsoaffiliation{AGH University of Science and Technology, Faculty of Physics and Applied Computer Science, 30-059 Kraków, Poland}

\begin{abstract}
We report on the pressure evolution of the electronic and crystal structures of the non-centrosymmetric antiferromagnet EuCoGe$_3$. Using a diamond anvil cell, we performed high pressure fluorescence detected near-edge x-ray absorption spectroscopy at the Eu  $L_3$, Co $K$, and Ge $K$ edges and synchrotron powder x-ray diffraction. In the Eu $L_3$ spectrum, both divalent and trivalent Eu peaks are observed from the lowest pressure measurement ($\sim$2 GPa). By increasing pressure, the relative intensity of the trivalent Eu peak increases, and an average Eu valence continuously increases from 2.2 at 2 GPa to 2.31 at $\sim$50 GPa. On the other hand, no discernible changes are observed in the Co $K$ and Ge $K$ spectra as a function of pressure. With the increase in pressure, lattice parameters continuously decrease without changing $I4mm$ symmetry. Our study revealed a robust divalent Eu state and an unchanged crystal symmetry of EuCoGe$_3$ against pressure.
\end{abstract}

\maketitle

%\textbf{Key Words} – EuCoGe$_3$, Eu valence, high pressure, and crystal structure.

\section{Introduction}
Intermetallic compounds with lanthanoids host various fascinating phenomena, such as heavy fermion behavior, spin/charge ordering, Kondo effect, and superconductivity, originating from an interplay of strongly correlated $4f$ electrons and itinerant conduction electrons \cite{Pfleiderer_RMP_2009, Si2010}.
A plethora of ternary lanthanoid transition metal silicides/germanides crystallize with the ThCr$_2$Si$_2$-type structure ($I4/mmm$) \cite{Just1996}, for instance, the first heavy fermion superconductor CeCu$_2$Si$_2$ \cite{Steglich1892} and the quantum critical Kondo lattice YbRh$_2$Si$_2$ \cite{Trovarelli2000, Gegenwart2002}. In isostructural europium-based silicides, Eu ions bear a divalent valence state Eu$^{2+}$ (4\textit{f}$^7$, $J$= 7/2) that favors an antiferromagnetic ground state \cite{Abd-Elmeguid1985,Hess1997,Mitsuda2012}. However, the energy difference between Eu$^{2+}$ and nonmagnetic Eu$^{3+}$ (4\textit{f}$^6$, $J$= 0) valence states is not very large \cite{Bauminger1973} and can be tuned by applying pressure and/or by chemical substitutions. Applying pressure or substituting smaller ions
tend to increase the antiferromagnetic transition temperature ($T_{\rm N}$), followed by a sudden disappearance of magnetic moments and a valence crossover at a critical pressure. Indeed, a pressure-induced Eu valence transition with a simultaneous collapse of antiferromagnetism was reported for Eu(Pd$_{0.8}$Au$_{0.2}$)$_2$Si$_2$ \cite{Abd-Elmeguid1985}, EuNi$_2$Si$_2$ \cite{Hess1997}, and EuRh$_2$Si$_2$ \cite{Mitsuda2012}, and a substitution-induced valence transition was found in EuNi$_2$(Ge$_{1-x}$Si$_x$)$_2$ \cite{Wada1999} and Eu(Pt$_{1-x}$Ni$_x$)$_2$Si$_2$ \cite{Mitsuda2007}. Due to the different ionic radii of Eu$^{2+}$ and Eu$^{3+}$ ions \cite{Shannon1976}, the Eu valence transition and the ground state properties in such systems are usually discussed in relation to the lattice volume. It has been established that compounds with a large unit cell volume possess an antiferromagnetic ground state with Eu$^{2+}$ ions, while materials with a small unit cell volume exhibit a nonmagnetic ground state with Eu$^{3+}$ ions\cite{Onuki2020, Honda2017}.

Contrary to rather extensive studies on the Eu-122 systems, much less attention has been given to ternary Eu-compounds crystallizing with the BaNiSn$_3$-type structure ($I4mm$) which is a close relative to the ThCr$_2$Si$_2$-type structure (see Fig. \ref{fig7} (b)). 
 Recently a series of europium transition metal silicides/germanides Eu$TX_3$ : \textit{T}= transition metal, \textit{X}=Si or Ge, with the BaNiSn$_3$-type structure \cite{Venturini1985} was reported to exhibit complex magnetic properties\cite{Maurya2014, Maurya2016, Fabreges2016, Bednarchuk_JAC_2015, Kakihana2017, Bauer2022, Matsumura2022} and atypical behavior under hydrostatic pressure\cite{Nakashima_JPSJ_2017, Nakamura2015}.
In this context, it is worth mentioning that pressure-induced superconductivity was discovered in a few Ce-based counterparts \cite{Kimura2005, Sugitani2006, Settai2007, Honda2010}. These compounds bear an unconventional character with a mixed singlet-triplet pairing caused by large anti-symmetric spin-orbit coupling in strongly correlated electron systems which lack an inversion symmetry in their crystal lattice\cite{Takimoto_JPSJ_2008, Takimoto_JPSJ_2009}.

In the crystallographic unit cell of Eu$TX_3$ systems, Eu atoms occupy the 2$a$ Wyckoff site, silicon/germanium atoms are located at two different Wyckoff positions 2$a$ and 4$b$, while transition metal atoms occupy the 2$a$ site \cite{Bednarchuk_JAC_2015}. Magnetic susceptibility measurements \cite{Bednarchuk_JAC_2_2015,Bednarchuk_APPA_2015, Kakihana2017} and Mössbauer spectroscopy \cite{Maurya2014, Maurya2016} revealed the presence of magnetic Eu$^{2+}$ ions in each of the investigated compounds. While all of them order antiferromagnetically (AFM) at similar temperatures, the magnetic structure formed by the localized Eu 4$f$ moments depends on the transition metal constituent. For example, in EuRhGe$_3$ the AFM order sets in at $T_{\rm N}$= 11.3 K and the Eu moments are confined in the \textit{ab} plane, while they are aligned along the $c$-axis in EuCoGe$_3$, EuNiGe$_3$ and EuIrGe$_3$ that order at $T_{\rm N}$ = 15.4, 13.5, and 12.3 K respectively \cite{Bednarchuk_JAC_2015,Bednarchuk_JAC_2_2015, Bednarchuk_APPA_2015}. Below $T_{\rm N}$, successive magnetic phase transitions were observed at $T'_{\rm N}$= 13.4 K in EuCoGe$_3$ and $T'_{\rm N}$= 7.5 K and $T$$^*_{\rm N}$= 5.0 K in EuIrGe$_3$ \cite{Bednarchuk_JAC_2015, Kakihana2017}. Very recently, EuIrGe$_3$ was studied by neutron and resonant x-ray diffraction and complex magnetic phase transitions from an incommensurate longitudinal sinusoidal structure below $T_{\rm N}$ to a cycloidal structure below $T'_{\rm N}$, then to a cycloidal structure rotated by 45\textdegree ~in-plane below $T^*_{\rm N}$ were revealed \cite{Matsumura2022}.

In order to check for possible valence changes and long-sought emergence of superconductivity in Eu-based materials, electric transport measurements were performed on EuCoGe$_3$, EuNiGe$_3$, EuRhGe$_3$, and EuIrGe$_3$ under pressure up to 8 GPa \cite{Uchima2014, Kakihana2017}. The magnetic transition temperatures $T_{\rm N}$ and $T'_{\rm N}$ were found to increase with increasing pressure and no sign of any other phase transition was observed. Similar results were obtained from pressure-dependent ac calorimetry in EuCoGe$_3$ up to 10.4 GPa, which additionally indicated a pressure-driven moderate effective mass enhancement \cite{Muthu2019}.

In this study, we performed high energy resolution fluorescence detected (HERFD) near-edge x-ray absorption spectroscopy and powder x-ray diffraction on EuCoGe$_3$ under pressure as high as 50 GPa.
By increasing pressure, the average Eu valence of EuCoGe$_3$ continuously increases from 2.2 at 2 GPa to 2.31 at $\sim$50 GPa, while no discernible changes are observed in the Co $K$ and Ge $K$ spectra as a function of pressure. Concurrently, the crystal lattice volume of EuCoGe$_3$ continuously decreases without changing the $I4mm$ symmetry.

\begin{figure*}[!t]
    \centering
    \includegraphics[scale=0.75]{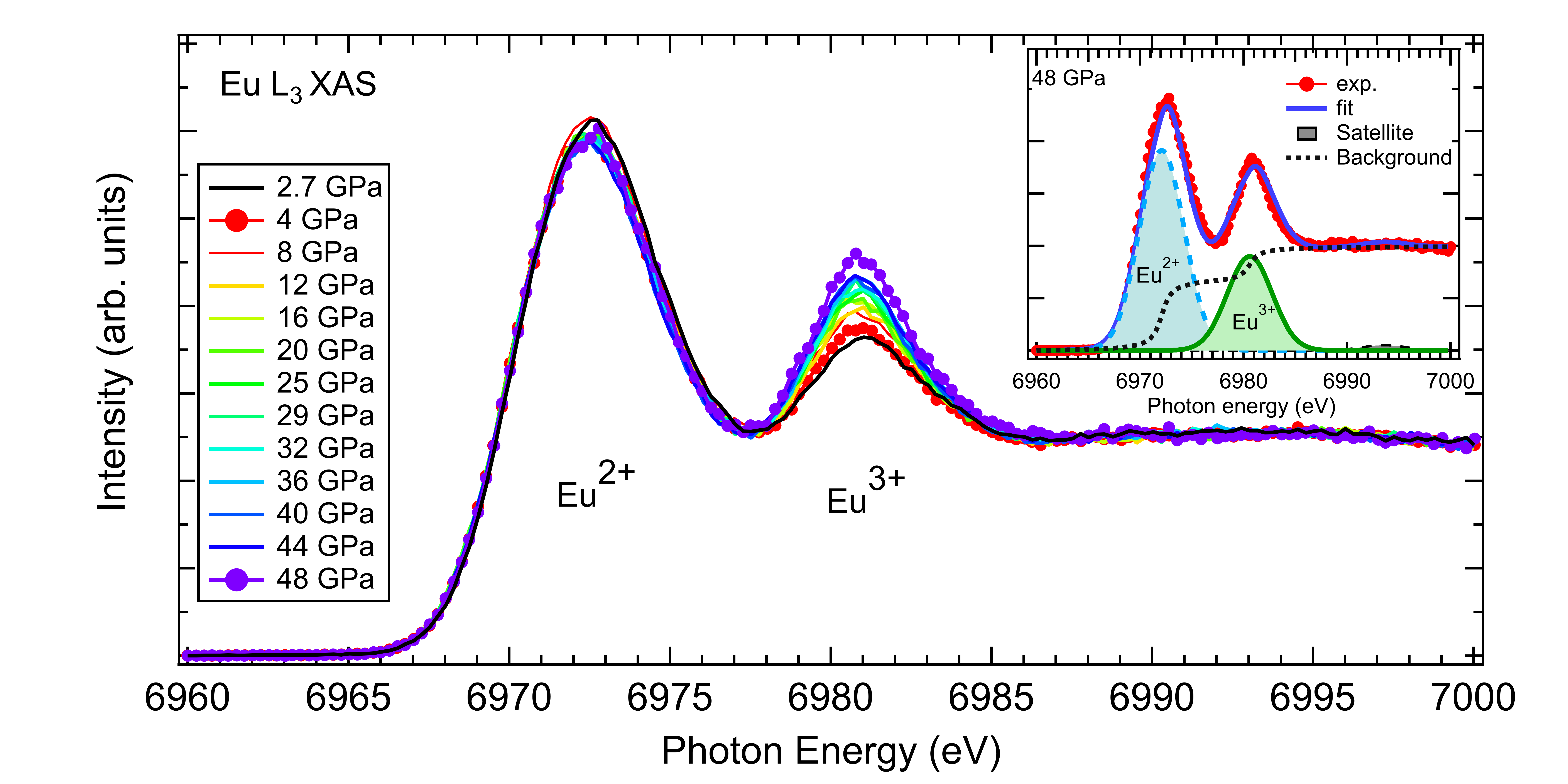}
    \caption{The Eu $L_3$ HERFD near-edge spectra of EuCoGe$_3$ at selected pressures. The spectra are normalized over the higher energy end after subtracting the constant background below the edge. Inset shows an example of fitting the Eu $L_3$ XAS spectrum at 48 GPa in order to extract the Eu$^{2+}$ and Eu$^{3+}$ components.}
   \label{fig1}
\end{figure*}

\section{Experiment}
Single crystals of EuCoGe$_3$ were grown from the In flux, as described elsewhere \cite{Bednarchuk_JAC_2015, Bednarchuk_APPA_2015, Kakihana2017}. The crystals were taken from the same batch that was studied in Ref. 20. Their high quality was proved by electrical resistivity and magnetic susceptibility measurements (see Ref. 20).

HERFD near-edge x-ray absorption spectroscopy (XAS) experiment was performed at the GALAXIES beamline of the SOLEIL synchrotron \cite{Rueff_2014, Ablett_2019}. The incident synchrotron beam was monochromatized using a Si(111) double-crystal monochromator followed by a Pd-coated spherical collimating mirror \cite{James_2021}. The HERFD near-edge XAS spectra were observed by varying photon energy across the Eu $L_3$, Co $K$, and Ge $K$ edges and recorded using a silicon drift detector. Each fluorescence line was selected by changing the Bragg angle of the single crystal analyzer: Eu $L_{\alpha1}$ (5846 eV), Co $K_{\beta1}$ (7649 eV), and Ge $K_{\alpha1}$ (9886 eV) with the Bragg angles of 77\textdegree~Ge(333), 84\textdegree ~Ge(444), and 74\textdegree ~Ge(555), respectively. HERFD method suppresses 2\textit{p} or 1\textit{s} core-hole lifetime broadening owing to the resonant inelastic x-ray scattering process. The sample was mounted in a diamond anvil cell (DAC) with Ne gas as a pressure medium and a ruby as a pressure indicator \cite{Mao_1986}. A high purity beryllium gasket was used through which the incident and fluorescence x-rays traverse to measure the Eu $L_{\alpha1}$, Co $K_{\beta1}$, and Ge $K_{\alpha1}$ emissions.  The pressure was applied by manually tightening a set of 4 screws on the DAC. For an accurate pressure calibration, the ruby fluorescence signal was measured before and after the XAS cycle at each pressure. 

The high-resolution x-ray diffraction (XRD) was carried out at the PSICHE beamline of SOLEIL synchrotron with a photon energy of 23 keV ($\lambda$ = 0.3738 \AA). The single crystal of EuCoGe$_3$ was gently crushed by pestle into powder. The powdered sample was mounted in a DAC with Ne gas as a pressure medium, and a piece of Au was loaded in the DAC together with the sample as a pressure reference material. The gasket was made of inox with a thickness of 27 $\mu$m and a sample space diameter of 150 $\mu$m.  Diamonds of 300 $\mu$m diameter culet size were used. The pressure was controlled by a membrane on the DAC and was determined by the Au equation of state.

\section{Results and Discussion}
\subsection{HERFD near-edge XAS results}

Fig. \ref{fig1} shows the pressure-dependent Eu $L_3$ XAS spectra of EuCoGe$_3$ measured at room temperature. The spectra were obtained by scanning the incident x-ray energy through the Eu $L_3$ absorption edge while recording the scattered intensity of the Eu $L_{\alpha1}$ fluorescence energy. The HERFD spectra resemble a standard XAS spectrum, though the spectral shape is sharper due to the absence of a deep 2$p$ core hole in the final state. The Eu $L_3$ HERFD spectra exhibit a prominent peak at 6972 eV and a broad peak centered at 6981 eV corresponding to Eu$^{2+}$ ($2p^64f^7 \rightarrow 2p^54f^7$ + $\epsilon d$($s$)  $\rightarrow 2p^63d^94f^7$) and Eu$^{3+}$ ($2p^64f^6  \rightarrow 2p^54f^6$ + $\epsilon d$($s$) $\rightarrow 2p^63d^94f^6$) components, respectively. Although the transition process does not directly include the 4$f$ states, the Eu$^{2+}$ and Eu$^{3+}$ peaks in the Eu $L_3$ HERFD spectra are well separated and they are sensitive to the change of the Eu valence due to strong Coulomb interaction between the 3$d$ core hole and the final state 4$f$ electron. A broad satellite peak around 20 eV above the main line is not evident in EuCoGe$_3$, in contrast to that observed in EuRhGe$_3$\cite{Utsumi_ES_2021}. The inset of Fig. \ref{fig1} shows the Eu XAS spectra fitting with three Gaussian functions corresponding to Eu$^{2+}$, Eu$^{3+}$ and the satellite peak, and two arctangent backgrounds for the Eu$^{2+}$ and Eu$^{3+}$ peaks. There is no signal of the quadrupolar 2$p$ - 4$f$ transition in the pre-edge region. The mean Eu valence was estimated using the formula $v = 2 + I^{3+}/(I^{2+} + I^{3+}$), where $I^{2+}$ and $I^{3+}$ denote the integrated spectral intensities of the Eu$^{2+}$ and Eu$^{3+}$ peaks respectively, extracted from the fitting analysis. Note that the intensity of the satellite peak is not included in the Eu valence estimation.

\begin{figure}[!t]
    \centering
    \includegraphics[scale=0.375]{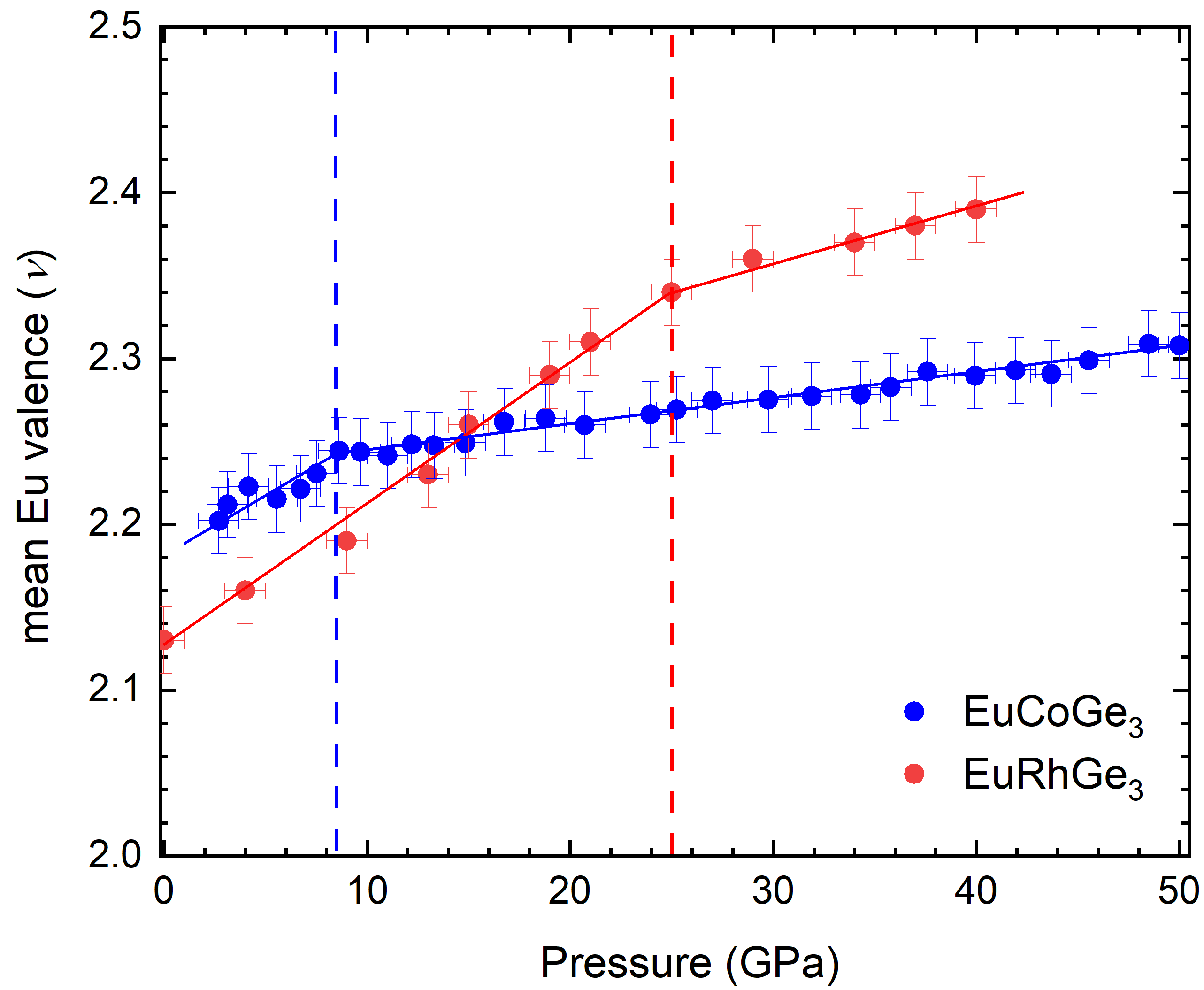}
    \caption{Pressure dependence of the Eu mean valence in EuCoGe$_3$ derived from the Eu $L_3$ HERFD spectra (blue symbols). For comparison, the data reported for EuRhGe$_3$ \cite{Utsumi_ES_2021} are also shown (red symbols).}
    \label{fig2}
\end{figure}

In Fig. \ref{fig2} we present the variation of the mean Eu valence with pressure in EuCoGe$_3$. The valence at 2.7~GPa (the lowest measurement pressure) is $v$ = 2.20 $\pm$ 0.02. With increasing pressure, $v$ linearly increases to 2.31 $\pm$ 0.02 at 50~GPa. For comparison, in Fig. \ref{fig2} we also show pressure variation of the mean Eu valence in EuRhGe$_3$ \cite{Utsumi_ES_2021}. No valence/phase transition was observed in the entire pressure range investigated. Compared to the latter compound, the pressure evolution of Eu valence in EuCoGe$_3$ is rather small. The rate of valence change $dv$/$dP$ in EuRhGe$_3$ is around 0.0065/GPa, while that in EuCoGe$_3$ is only 0.0023/GPa. Similarly a stable Eu$^{2+}$ valence state is relatively uncommon for Eu-compounds. The origin of the kinks in the Eu valence changes is unclear, though they are not artefacts created by the pressure medium since both were measured with a Ne gas pressure medium. For the most intensively studied ternaries with the ThCr$_2$Si$_2$-type structure, a pressure-induced valence transition from Eu$^{2+}$ to almost Eu$^{3+}$, usually occurs in the range 4-5 GPa. \cite{Honda2017}\cite{Onuki_PM_2017}

\begin{figure*}[!t]
    \centering
    \includegraphics[scale=0.5]{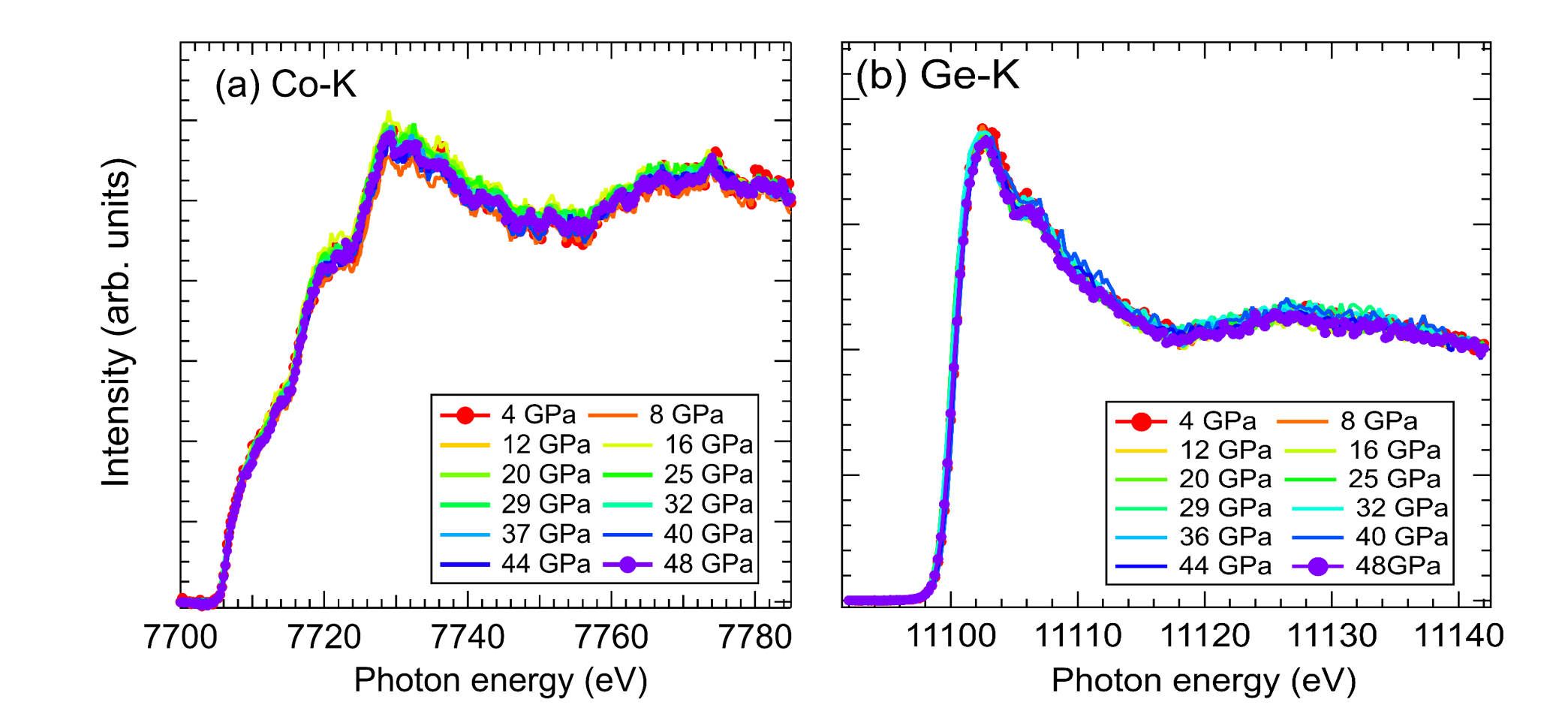}
    \caption{HERFD near-edge XAS spectra of EuCoGe$_3$  taken at (a) Co $K$ edge and (b) Ge $K$ edge normalized on the higher energy end.}
    \label{fig3}
\end{figure*}

In Fig \ref{fig3}, the HERFD spectra at the Co $K$ and Ge $K$ edges are shown. After subtracting a constant background below the edges, the spectra were normalized over the higher energy end. The Co $K$ edge spectra show pre-edge shoulder structures at 7710~eV and 7720~eV and the main peak at 7729~eV which corresponds to the Co 1$\textit{s}  \rightarrow4\textit{p}$ dipolar transition. The pre-edge structures were reported for Co foil \cite{Hlil1996} and Co-bearing oxides \cite{Vanko2006}. The pre-edge shoulder at 7710~eV can be attributed to the Co 1$s \rightarrow 3d$ direct quadrupolar transition and the dipolar transition to $d-p$ hybridised state \cite{Vanko2006, Hlil1996}. The following pre-edge structure at 7720~eV could be due to a shakedown process of ligand to metal charge transfer \cite{Kim1997}. The broad satellite peak far above the edge centred at 7775 eV may originate from extended x-ray absorption fine structure oscillations. The Ge $K$ edge spectra have a main peak at 11102~eV and a shoulder peak at 11106~eV. The spectral shape is similar to the Ge $K$ XAS spectrum of CeCoGe$_3$ which is isostructural to EuCoGe$_3$. Following the interpretation of Ge $K$ XAS spectrum of CeCoGe$_3$ \cite{Rogalev2021}, the prominent peak at 11102~eV and the shoulder peak at 11106~eV are considered to originate from Ge atoms in the Wyckoff positions $4b$ and $2a$, respectively. Within experimental resolution, neither Co $K$ nor Ge $K$ edge spectra show any remarkable changes with pressure. In order to elucidate a slight change by pressure, we also observed the Co $K_{\beta}$ and Ge $K_{\alpha}$ x-ray emission spectra, though no reasonable changes have been detected (see appendix). The results indicate that the increase of the mean Eu valence under pressure is due to intra-atomic charge transfer from Eu $4f$ to 5$d$, and negligible contributions from Ge and Co ions.

\subsection{XRD results}
\begin{figure*}[!t]
\centering
\includegraphics[scale=0.45]{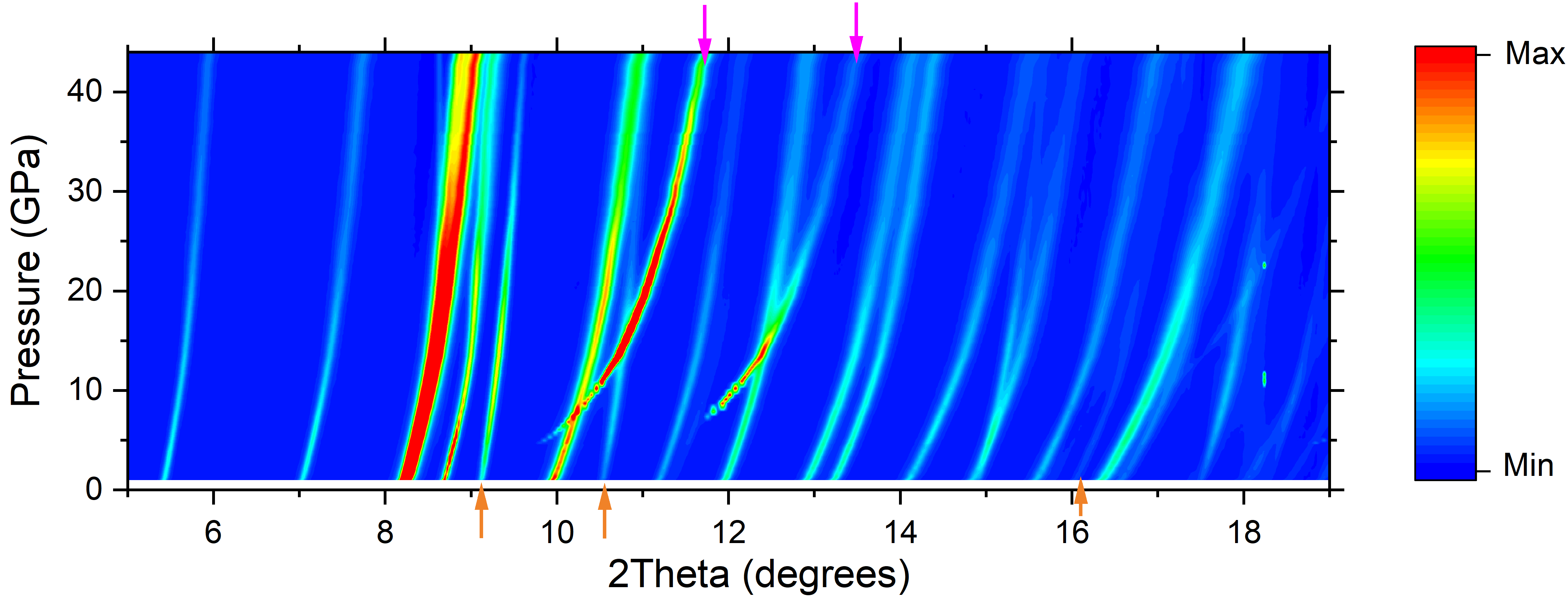}
\caption{Contour map of synchrotron x-ray diffraction intensities collected in
the pressure range 1 – 45~GPa, for EuCoGe$_3$ (main phase), gold (standard
material), and neon (pressure medium). The magenta arrows (top) and the yellow arrows (bottom) of the contour map represent the main peak positions of neon and gold respectively.}
\label{fig4}
\end{figure*}

Synchrotron powder XRD of EuCoGe$_3$ was performed as a function of pressure at room temperature. The contour map of diffraction intensities in the pressure range from 1 to 45 GPa is shown in Fig \ref{fig4}. In order to highlight the change with pressure, the diffraction intensities are plotted only in the 2$\theta$ range 5\textdegree~ to 19\textdegree.  The pressure evolution of Bragg peak positions indicates three different phases, EuCoGe$_3$ (main phase), gold, and neon as a hydrostatic pressure medium. Neon is known to crystallize around 4.8~GPa at room temperature \cite{Finger1981}. A sign of non-hydrostaticity in neon appears above 15~GPa, though its pressure gradient stays very small up to 50~GPa \cite{Meng1993, Klotz_IOP_2009}. The fast-changing peaks at 2$\theta$ = 10\textdegree ~and 12\textdegree ~ in the contour map correspond to crystalized neon. No emergence or disappearance of Bragg peaks was observed that could be related to any symmetry change in the investigated pressure range. This result proves the absence of any structural phase transition in EuCoGe$_3$ up to 45~GPa. In order to extract more detailed information on the crystal structure of EuCoGe$_3$, Rietveld refinement of the XRD data was performed by using the Profex programme.\cite{Profex}

\begin{figure}[!htb]
    \centering
    \includegraphics[scale=0.515]{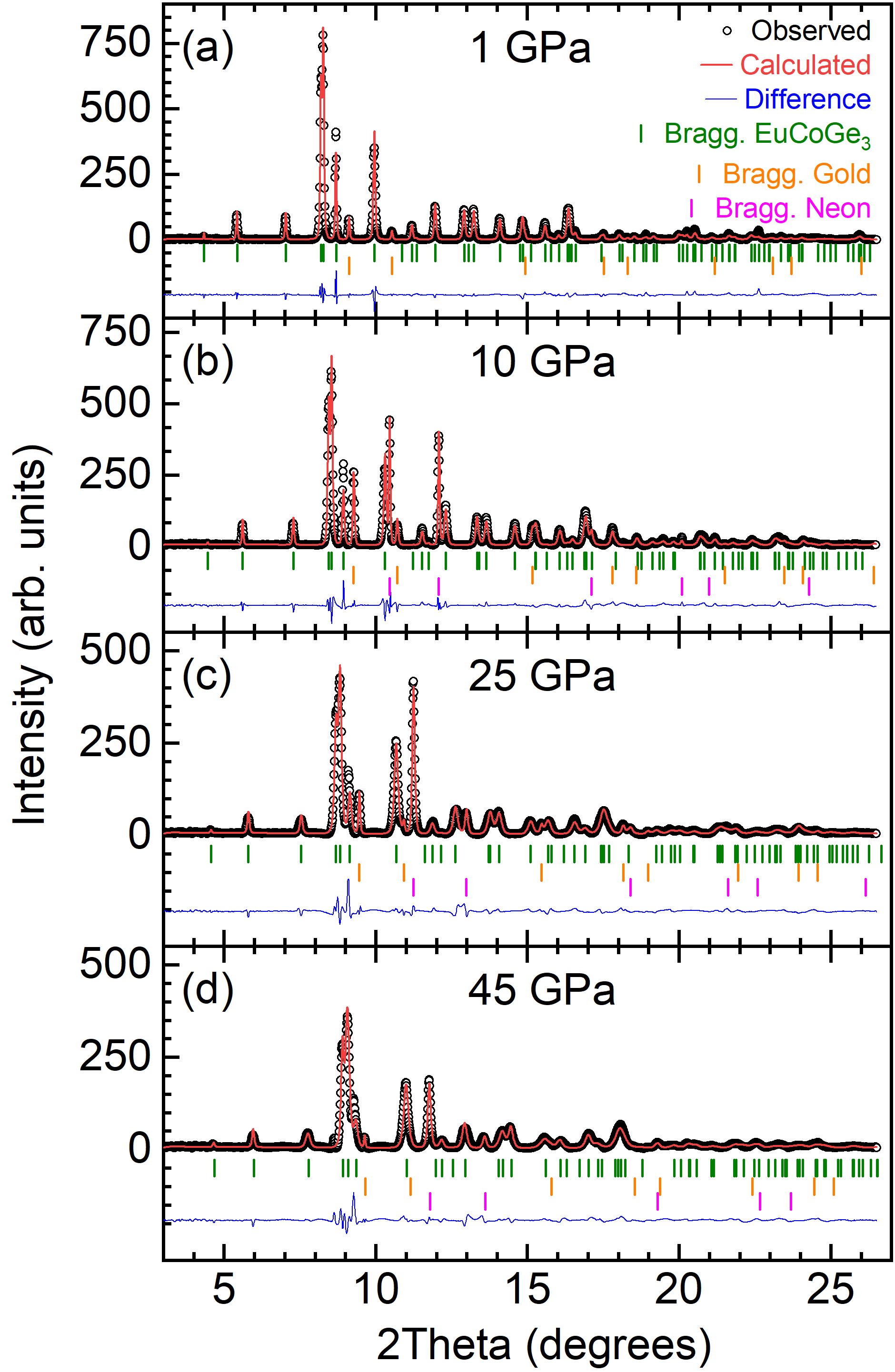}
    \caption{Integrated synchrotron XRD patterns of EuCoGe$_3$ with the results of refinement at (a) 1~GPa, (b) 10~GPa, (c) 25~GPa, and (d) 45~GPa. The vertical bars indicate Bragg peak positions of EuCoGe$_3$ (green), gold (orange), and neon (magenta).}
    \label{fig5}
\end{figure}

\begin{figure*}[!htb]
    \centering
    \includegraphics[scale=0.6]{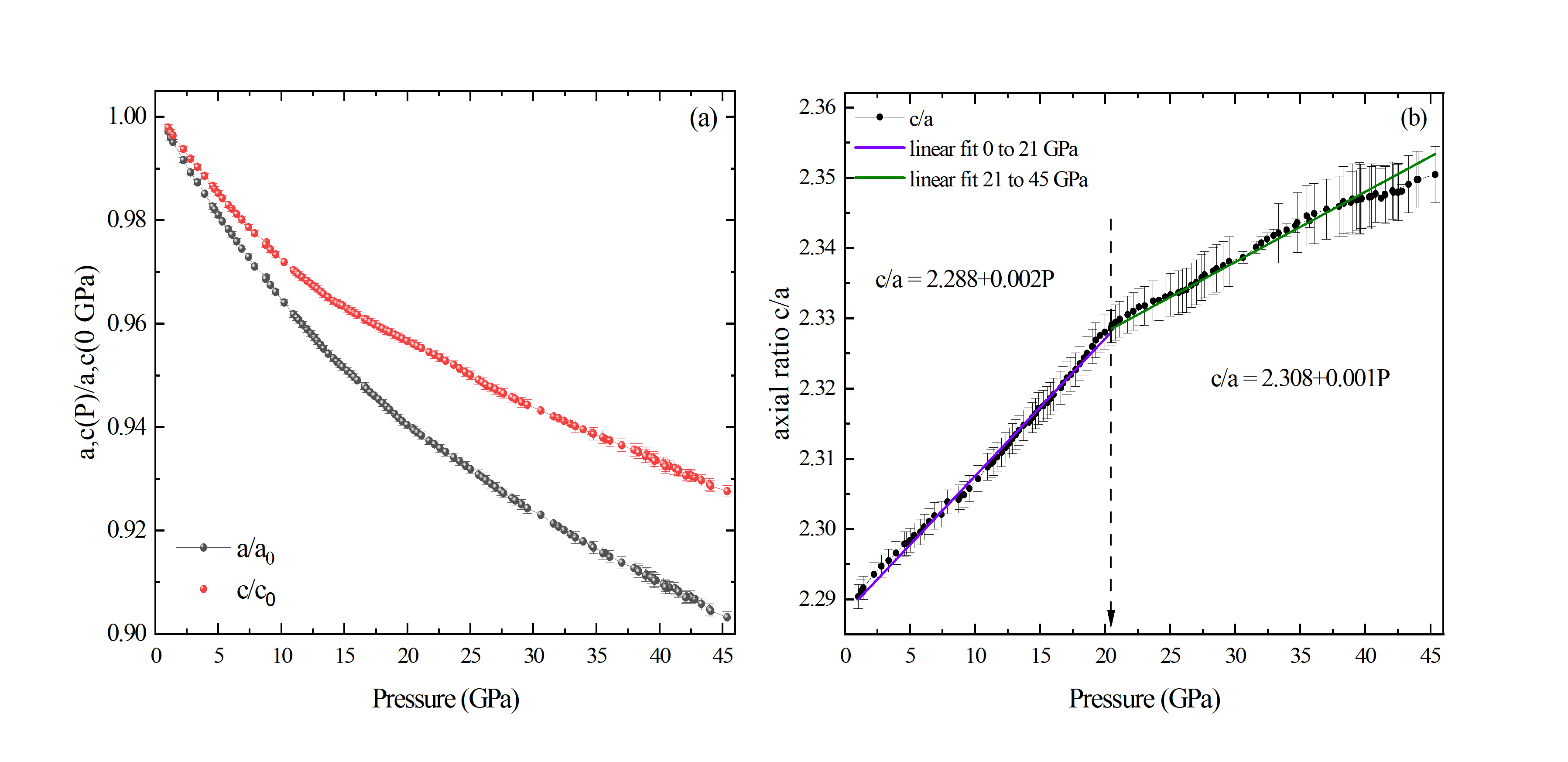}
    \caption{(a) Relative pressure variations of the lattice parameters of EuCoGe$_3$ with respect to the values at ambient pressure. (b) Pressure dependence of the ratio c/a. Straight lines emphasize linear behaviour.}
    \label{fig6}
\end{figure*}

In Fig. \ref{fig5}, the XRD patterns at 1, 10, 25, and 45 GPa are presented along with the results of Rietveld refinements (See also Table 1). The analysis confirmed that EuCoGe$_3$ keeps the same crystal symmetry ($I4mm$) up to 45~GPa. Fig. \ref{fig6} (a) shows the relative changes in both the $a$- and $c$-lattice parameters with respect to the values at ambient pressure : $a$= 4.3191(3)~\AA~and $c$= 9.8847(15)~{\AA} (taken from Ref \cite{Bednarchuk_JAC_2015}). In both directions, a smooth contraction with pressure (\textit{P}) was found, however, the change along the $a$-axis is larger than along the $c$-axis. The ratio $c/a$ is plotted in Fig. \ref{fig6} (b) as a function of pressure. It linearly increases with increasing pressure up to 21 GPa with a rate 2.288+0.002$\textit{P}$ and then continues to increase nearly linearly with a slope 2.308+0.001$\textit{P}$. The change of the $c/a$ increase rate around 20 GPa can be attributed to the anisotropic compression of $a$- and $c$-axis. A small deviation from the linear behavior above 40 GPa might be due to the non-hydrostatic pressure effect in high-pressure region.

Compared to the axial ratio of various ternary Eu-compounds in the ThCr$_2$Si$_2$-type structure ($I4/mmm$) which has a centrosymmetry \cite{Onuki2020}, that of EuCoGe$_3$ is relatively small. In the case of $AB_2X_2$ stoichiometric compounds ($A$: rare-earth or alkali earth, $B$: transition metal, and $X$: 14-15 group elements) with the ThCr$_2$Si$_2$-type structure, the structure can be disassemble to $B_2X_2$-layers and $A^{2+}$ ions. The $B_2X_2$-layers are constructed from $BX_4$ tetrahedrons that have covalent $B-X$ bonding and weak $B-B$ metal-metal bonding. The interrelationship between intra- and inter-layer bonding distances, namely the $c/a$ ratio, and its relation to the physical properties in $AB_2X_2$ compounds have been proposed \cite{Hoffmann1985, Johrendt1997, Huhnt1998, Reehuis1998}. Although the correlation between lattice parameters and physical properties may exist in $ABX_3$ compounds in the BaNiSn$_3$-type structure, the bonding nature can differ from the ThCr$_2$Si$_2$-type compounds. In $ABX_3$ compounds, transition metal and $X$ atoms form square pyramids with the apex of $B/X$ atom pointing alternately up and down in the $BX_3^{-2}$ layers \cite{Li1986}. The metal-metal $B-B$ distance in the $BX_3^{-2}$ layer corresponds to the $a$-lattice constant and is relatively larger than that in the ThCr$_2$Si$_2$-type compounds. Furthermore, the interlayer distance is defined by the $B-X$ bonding. Here, we only refer to the structural differences between the ThCr$_2$Si$_2$-type and the BaNiSn$_3$-type compounds. In order to understand the relation between structure and properties in $ABX_3$ compounds, further systematic studies of $ABX_3$ series and chemical bonding analyses are necessary.

\begin{table*}[t!]
\centering
\begin{tabular}{||c c c c c||}
 \hline
 Pressure & 1 GPa & 10 GPa & 25 GPa & 45 GPa \\ [1.2ex]
 \hline\hline
   Lattice constants \\

a  (\AA) & 4.308 (4) & 4.164 (2) & 4.019 (6) & 3.901(1)\\
c  (\AA) & 9.867 (1) & 9.606 (9) & 9.379 (2) & 9.167 (1)\\
 V (\AA$^3$) & 183.156 (1) & 166.573 (1) & 151.541 (4) & 139.472 (9)\\
Refinement parameters\\
R$_{WP}$ & 17.84 & 16.12 & 17.69 & 21.16\\
R$_{exp}$ & 21.87 & 19.74 & 20.68 & 22.04\\
$\chi^2$& 0.674 & 0.6669 & 0.7317 & 0.9231\\
GOF& 0.821 & 0.8166 & 0.8554 & 0.9601\\  [1.2ex]
\hline
\hline
\end{tabular}
\caption{Lattice constants, unit cell volume and refinement parameters of EuCoGe$_3$ at selected pressures. Where $R_{WP}$ is the weighted profile R factor, $R_{exp}$ is the expected R factor and GOF is the goodness of fitting.}
\label{table:1}
\end{table*}

\begin{figure*}[!htb]
    \centering
    \includegraphics[scale=0.6]{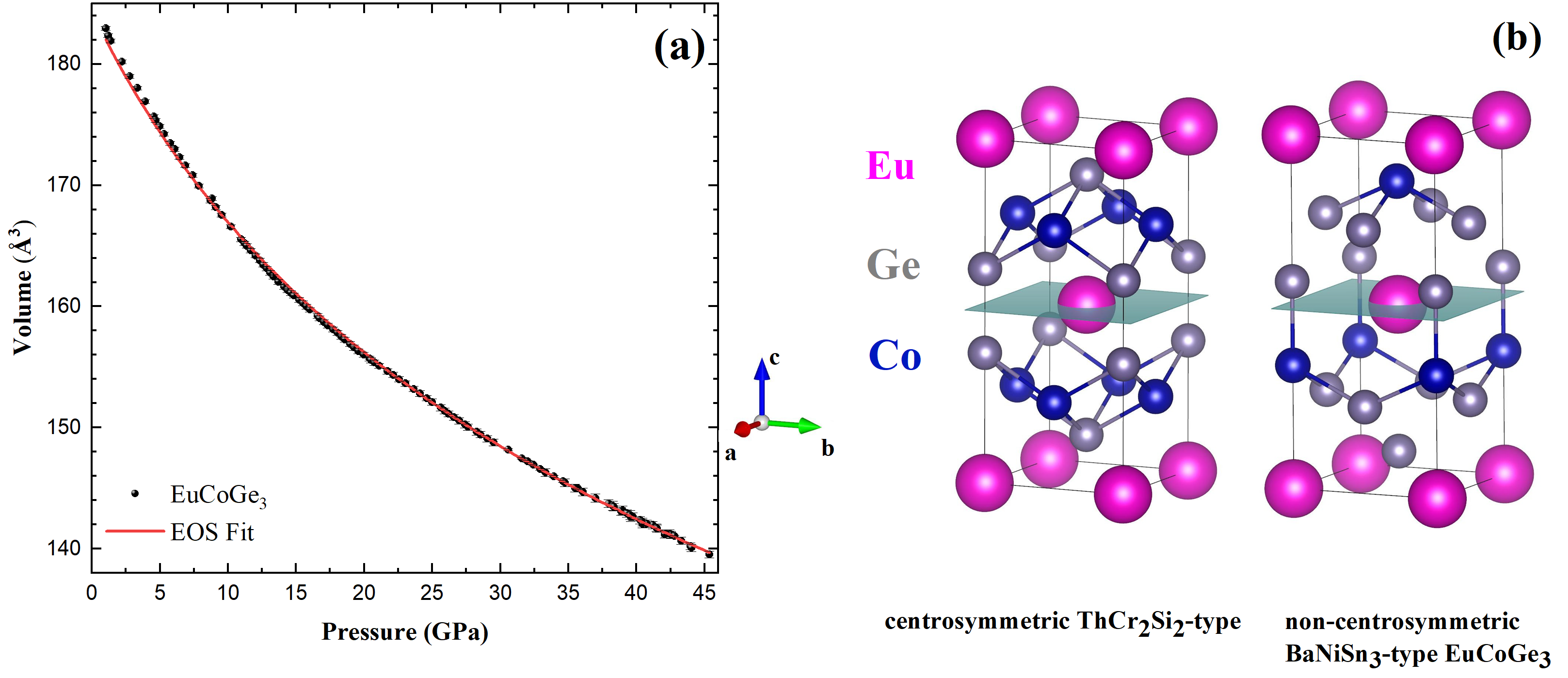}
    \caption{ (a) Pressure evolution of the unit cell volume of EuCoGe$_3$ with the result of EOS fitting. The error bar is the range of symbol size. (b) The centrosymmetric ThCr$_2$Si$_2$-type crystal structure and the non-centrosymmetric BaNiSn$_3$-type crystal structure of EuCoGe$_3$ showing Co-Ge bonds drawn by using VESTA \cite{Momma2011}. The inversion symmetry is broken with respect to the (002) plane.}
    \label{fig7}
\end{figure*}

The pressure evolution of the unit cell volume also shows a smooth contraction as shown in Fig. \ref{fig7}. The contraction of the unit cell volume can be approximated a linear decrease up to $\sim$10 GPa, which almost coincides the kink in the pressure change of the mean Eu valence. The linear increases of the mean Eu valence and $T_{\rm N}$ by applying pressure \cite{Kakihana2017, Muthu2019} might be related to the linear contraction of the unit cell volume, at least up to $\sim$10 GPa (See Supplemental Material Fig. S1 \cite{Supplement}). However, the pressure change of the Eu valence does not seem to be proportional to the pressure change of the unit cell volume in higher pressure region. As mentioned earlier, the ground state properties of the Eu-122 system have been discussed in relation to the unit cell volume \cite{Onuki2020, Honda2017}. Such a tendency was also reported in the Eu-113 system. With an increase of Ge substitution in EuNi(Si$_{1-x}$Ge$_x$)$_3$, the unit cell volume showed a monotonous increase, while $T_{\rm N}$ decreased monotonously, indicating its strong connection to the volume change \cite{Uchima2014b}. On the other hand, the change in $T_{\rm N}$ in Eu$T$Ge$_3$ by transition metal substitution did not show a proportional change with the unit cell volume \cite{Bednarchuk_JAC_2015}. Although it has to be taken into account that the chemical pressure and applying external pressure do not always lead to the same results, the Eu-113 systems do not likely follow the same unit cell volume regime as the Eu-122 systems.

We performed equation of state (EOS) fitting for EuCoGe$_3$ by using the EOSFit7c software\cite{EosFit}. We used the third-order Birch Murnaghan equation below for EOS fitting \cite{Birch1947}:

%\[P(V)= \frac{3B_0}{2} [(V_0/V)^{7/3} - (V_0/V)^{5/3} ](1+\frac{3}{4}(B_0'-4)[(V_0/V)^{7/3}  -1])\]

\begin{multline*}
	P(V)= \frac{3B_0}{2} [(V_0/V)^{7/3} - (V_0/V)^{5/3} ]\\ 
	{\times} \{1+\frac{3}{4}(B_0'-4)[(V_0/V)^{2/3}  -1]\}
	\end{multline*}

Where,
\[B_0 = -V (\frac{\partial P}{\partial V})_{P=0} \hspace{5pt} \& \hspace{5pt} B^{\prime}_0 = (\frac{\partial B}{\partial P})_{P=0}\]

Here $B_0$ and $B^{\prime}_0$ denote a bulk modulus at 0~GPa and the first pressure derivative of the bulk modulus, respectively. The result of EOS fitting is presented in Fig. \ref{fig7}. The value of $V_0$= 184.39 $\AA$$^3$ was taken from Ref. \cite{Bednarchuk_JAC_2015}. The obtained values of  $B_0$ and $B^{\prime}_0$ are 75.6 $\pm$ 0.3 GPa and 5.58 $\pm$ 0.04, respectively (See Supplemental Material Fig. S2 \cite{Supplement}). We also performed refinement for the gold reference material and neon pressure medium and extracted the pressure-dependent unit cell volumes. From the obtained unit cell volume of gold, the pressure was determined by using third-order Birch Murnaghan equation with $V_0$=67.847 $\AA^3$,  $B_0$=167 GPa, and $B^{\prime}_0$=5.5 taken from Ref. \cite{EosGold_1984}. %EOS fitting parameters of these materials are in good agreement with the former studies \cite{EosGold_1984, Yoneda_IOP_2017, Hemley_PRB_1989}. For gold, the obtained values of  $V_0$,   and $B^{\prime}_0$ are 67.917(4) Å$^3$, 173 ± 5 GPa, 4.5 ± 0.01 respectively.  
For the EOS fitting of the unit cell volume of neon (See Supplemental Material Fig. S3 \cite{Supplement}), we used $V_0$ = 88.967 $\AA^3$  from Ref. \cite{Hemley_PRB_1989} and obtained $B_0$= 1.15 $\pm$ 0.06 GPa and $B^{\prime}_0$= 9.0 $\pm$ 0.3, that are in good agreement with earlier studies \cite{Finger1981, Hemley_PRB_1989}.

\section{Conclusion}
We studied the electronic and crystal structure of non-centrosymmetric EuCoGe$_3$ by HERFD near-edge XAS and synchrotron powder XRD as a function of pressure. By applying pressure, the intensity of the Eu$^{3+}$ peak in the Eu $L_3$ XAS spectra slightly increases relative to the Eu$^{2+}$ peak. The mean Eu valence increases from 2.2 at 2.7~GPa to 2.31 at 50~GPa without a first-order valence transition. Compared to EuRhGe$_3$, the pressure variation of the Eu valence is relatively small in EuCoGe$_3$. The XAS spectra at the Ge $K$ and Co $K$ edges show no discernible changes against pressure. This indicates that the pressure evolution of the mean Eu valence is due to charge transfer from Eu $4f$ to 5$d$, with no contribution from Ge and Co ions.

The powder XRD experiment revealed a continuous compression of the lattice volume without changing structural symmetry within the investigated pressure range. The compressibility of the lattice constant along the $a$-axis is larger than that along the $c$-axis. From EOS fitting of the unit cell volume, we obtained the bulk modulus and the pressure derivative of the bulk modulus of EuCoGe$_3$. The contraction of the unit cell volume may contribute to the increase of the Eu valence, as our experimental results seem to show a relation between the mean Eu valence and the pressure evolution of the unit cell volume in EuCoGe$_3$ up to $\sim$10 GPa. However, the pressure change of the Eu valence does not seem to be proportional to the pressure change of the unit cell volume in higher pressure region.

\section{Acknowledgments}
We would like to thank Dominique Prieur for his skillful technical assistance. The HERFD near-edge XAS experiment was performed under the approval of the GALAXIES beamline station (proposal No. 20191156) and powder XRD experiment was performed under the approval of the PSICHE beamline station (proposal No. 20210410). The work has been supported by the Croatian Science Foundation under project number UIP-2019-04-2154. This work has been supported in part by the Croatian Science Foundation under the Project No. IP-2020-02-9666. The research leading to this result has been supported by the project CALIPSOplus under the Grant Agreement 730872 from the EU Framework Program for Research and Innovation HORIZON 2020. N. S. D acknowledges financing from the Croatian Science Foundation under the "Young Researchers' Career Development Project": project number DOK-2018-09-9906. C. M. N. Kumar acknowledges support of project Cryogenic Centre at the Institute of Physics - KaCIF- (Grant No. KK.01.1.1.02.0012), co-financed by the Croatian Government and the European Union through the European Regional Development Fund - Competitiveness and Cohesion Operational Programme.

	\global\long\def\appendixname{APPENDIX}
	\section{APPENDIX}

\appendix
	
	%\section{X-ray Emission Spectroscopy}
	%\setcounter{figure}{0}
%\renewcommand{\thefigure}{A\arabic{figure}}

\begin{figure*}[!htb]
    \centering
    \includegraphics[scale=0.45]{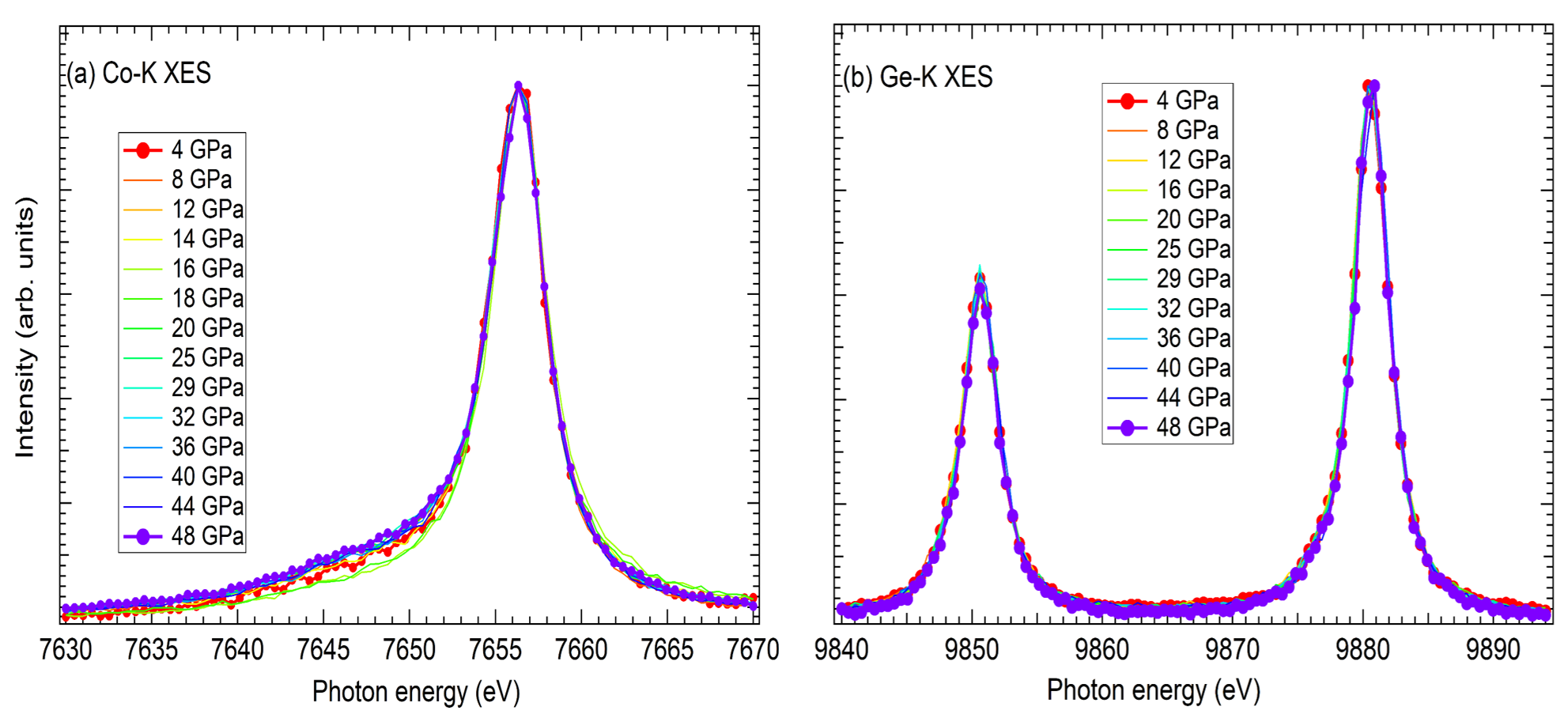}
    \caption{(a) X-ray emission spectra (XES) of Co \textit {K$_{\beta}$}, spectra is normalized on the emission peak K$_{\beta 1,3}$ intensity and (b) Ge \textit {K$_\alpha$} line, spectra is normalized at the emission peak intensity \textit {K$_{\alpha 1 }$}.}
  \label{fig8}
\end{figure*}

In Fig. 8, we present (a) Co \textit {K$_\beta$}  and (b) Ge \textit {K$_\alpha$} emission spectra of EuCoGe$_3$ as a function of pressure. The Co \textit {K$_\beta$} and Ge \textit {K$_\alpha$} emission spectra were recorded at incident photon energies of 8209 and 11603 eV, respectively. The Co \textit {K$_\beta$}  emission spectrum represents the main peak located at 7656~eV corresponding to the K$_{\beta 1,3}$ line. The satellite peak between 7645~eV and 7650 eV corresponds to the K$\beta^\prime$ line. The intensity of the satellite peak changes slightly but no systematic change as a function of pressure is observed. For emission along Ge \textit {K$_\alpha$}, we have two peaks, the first peak at 9850~eV which corresponds to \textit {K$_{\alpha 2}$} and the second peak at 30~eV higher at 9880~eV, which corresponds to \textit {K${_{\alpha1}}$}. Similar to XAS spectra no considerable changes were observed in spectra within experimental resolution as a function of pressure.

\clearpage
\pagebreak

\bibliographystyle{apsrev}
	\bibliography{EuCoGe3_reference}

\begin{thebibliography}{61}
\expandafter\ifx\csname natexlab\endcsname\relax\def\natexlab#1{#1}\fi
\expandafter\ifx\csname bibnamefont\endcsname\relax
  \def\bibnamefont#1{#1}\fi
\expandafter\ifx\csname bibfnamefont\endcsname\relax
  \def\bibfnamefont#1{#1}\fi
\expandafter\ifx\csname citenamefont\endcsname\relax
  \def\citenamefont#1{#1}\fi
\expandafter\ifx\csname url\endcsname\relax
  \def\url#1{\texttt{#1}}\fi
\expandafter\ifx\csname urlprefix\endcsname\relax\def\urlprefix{URL }\fi
\providecommand{\bibinfo}[2]{#2}
\providecommand{\eprint}[2][]{\url{#2}}

\bibitem[{\citenamefont{Pfleiderer}(2009)}]{Pfleiderer_RMP_2009}
\bibinfo{author}{\bibfnamefont{C.}~\bibnamefont{Pfleiderer}},
  \bibinfo{journal}{Rev. Mod. Phys.} \textbf{\bibinfo{volume}{81}},
  \bibinfo{pages}{1551} (\bibinfo{year}{2009}),
  \urlprefix\url{https://link.aps.org/doi/10.1103/RevModPhys.81.1551}.

\bibitem[{\citenamefont{Si and Steglich}(2010)}]{Si2010}
\bibinfo{author}{\bibfnamefont{Q.}~\bibnamefont{Si}} \bibnamefont{and}
  \bibinfo{author}{\bibfnamefont{F.}~\bibnamefont{Steglich}},
  \bibinfo{journal}{Science} \textbf{\bibinfo{volume}{329}},
  \bibinfo{pages}{1161} (\bibinfo{year}{2010}),
  \eprint{https://www.science.org/doi/pdf/10.1126/science.1191195},
  \urlprefix\url{https://www.science.org/doi/abs/10.1126/science.1191195}.

\bibitem[{\citenamefont{Just and Paufler}(1996)}]{Just1996}
\bibinfo{author}{\bibfnamefont{G.}~\bibnamefont{Just}} \bibnamefont{and}
  \bibinfo{author}{\bibfnamefont{P.}~\bibnamefont{Paufler}},
  \bibinfo{journal}{Journal of Alloys and Compounds}
  \textbf{\bibinfo{volume}{232}}, \bibinfo{pages}{1} (\bibinfo{year}{1996}),
  ISSN \bibinfo{issn}{0925-8388},
  \urlprefix\url{https://www.sciencedirect.com/science/article/pii/0925838895019391}.

\bibitem[{\citenamefont{Steglich et~al.}(1979)\citenamefont{Steglich, Aarts,
  Bredl, Lieke, Meschede, Franz, and Sch\"afer}}]{Steglich1892}
\bibinfo{author}{\bibfnamefont{F.}~\bibnamefont{Steglich}},
  \bibinfo{author}{\bibfnamefont{J.}~\bibnamefont{Aarts}},
  \bibinfo{author}{\bibfnamefont{C.~D.} \bibnamefont{Bredl}},
  \bibinfo{author}{\bibfnamefont{W.}~\bibnamefont{Lieke}},
  \bibinfo{author}{\bibfnamefont{D.}~\bibnamefont{Meschede}},
  \bibinfo{author}{\bibfnamefont{W.}~\bibnamefont{Franz}}, \bibnamefont{and}
  \bibinfo{author}{\bibfnamefont{H.}~\bibnamefont{Sch\"afer}},
  \bibinfo{journal}{Phys. Rev. Lett.} \textbf{\bibinfo{volume}{43}},
  \bibinfo{pages}{1892} (\bibinfo{year}{1979}),
  \urlprefix\url{https://link.aps.org/doi/10.1103/PhysRevLett.43.1892}.

\bibitem[{\citenamefont{Trovarelli et~al.}(2000)\citenamefont{Trovarelli,
  Geibel, Mederle, Langhammer, Grosche, Gegenwart, Lang, Sparn, and
  Steglich}}]{Trovarelli2000}
\bibinfo{author}{\bibfnamefont{O.}~\bibnamefont{Trovarelli}},
  \bibinfo{author}{\bibfnamefont{C.}~\bibnamefont{Geibel}},
  \bibinfo{author}{\bibfnamefont{S.}~\bibnamefont{Mederle}},
  \bibinfo{author}{\bibfnamefont{C.}~\bibnamefont{Langhammer}},
  \bibinfo{author}{\bibfnamefont{F.~M.} \bibnamefont{Grosche}},
  \bibinfo{author}{\bibfnamefont{P.}~\bibnamefont{Gegenwart}},
  \bibinfo{author}{\bibfnamefont{M.}~\bibnamefont{Lang}},
  \bibinfo{author}{\bibfnamefont{G.}~\bibnamefont{Sparn}}, \bibnamefont{and}
  \bibinfo{author}{\bibfnamefont{F.}~\bibnamefont{Steglich}},
  \bibinfo{journal}{Phys. Rev. Lett.} \textbf{\bibinfo{volume}{85}},
  \bibinfo{pages}{626} (\bibinfo{year}{2000}),
  \urlprefix\url{https://link.aps.org/doi/10.1103/PhysRevLett.85.626}.

\bibitem[{\citenamefont{Gegenwart et~al.}(2002)\citenamefont{Gegenwart,
  Custers, Geibel, Neumaier, Tayama, Tenya, Trovarelli, and
  Steglich}}]{Gegenwart2002}
\bibinfo{author}{\bibfnamefont{P.}~\bibnamefont{Gegenwart}},
  \bibinfo{author}{\bibfnamefont{J.}~\bibnamefont{Custers}},
  \bibinfo{author}{\bibfnamefont{C.}~\bibnamefont{Geibel}},
  \bibinfo{author}{\bibfnamefont{K.}~\bibnamefont{Neumaier}},
  \bibinfo{author}{\bibfnamefont{T.}~\bibnamefont{Tayama}},
  \bibinfo{author}{\bibfnamefont{K.}~\bibnamefont{Tenya}},
  \bibinfo{author}{\bibfnamefont{O.}~\bibnamefont{Trovarelli}},
  \bibnamefont{and} \bibinfo{author}{\bibfnamefont{F.}~\bibnamefont{Steglich}},
  \bibinfo{journal}{Phys. Rev. Lett.} \textbf{\bibinfo{volume}{89}},
  \bibinfo{pages}{056402} (\bibinfo{year}{2002}),
  \urlprefix\url{https://link.aps.org/doi/10.1103/PhysRevLett.89.056402}.

\bibitem[{\citenamefont{Abd-Elmeguid et~al.}(1985)\citenamefont{Abd-Elmeguid,
  Sauer, and Zinn}}]{Abd-Elmeguid1985}
\bibinfo{author}{\bibfnamefont{M.~M.} \bibnamefont{Abd-Elmeguid}},
  \bibinfo{author}{\bibfnamefont{C.}~\bibnamefont{Sauer}}, \bibnamefont{and}
  \bibinfo{author}{\bibfnamefont{W.}~\bibnamefont{Zinn}},
  \bibinfo{journal}{Phys. Rev. Lett.} \textbf{\bibinfo{volume}{55}},
  \bibinfo{pages}{2467} (\bibinfo{year}{1985}),
  \urlprefix\url{https://link.aps.org/doi/10.1103/PhysRevLett.55.2467}.

\bibitem[{\citenamefont{Hesse et~al.}(1997)\citenamefont{Hesse, Lübbers,
  Winzenick, Neuling, and Wortmann}}]{Hess1997}
\bibinfo{author}{\bibfnamefont{H.-J.} \bibnamefont{Hesse}},
  \bibinfo{author}{\bibfnamefont{R.}~\bibnamefont{Lübbers}},
  \bibinfo{author}{\bibfnamefont{M.}~\bibnamefont{Winzenick}},
  \bibinfo{author}{\bibfnamefont{H.}~\bibnamefont{Neuling}}, \bibnamefont{and}
  \bibinfo{author}{\bibfnamefont{G.}~\bibnamefont{Wortmann}},
  \bibinfo{journal}{Journal of Alloys and Compounds}
  \textbf{\bibinfo{volume}{246}}, \bibinfo{pages}{220} (\bibinfo{year}{1997}),
  ISSN \bibinfo{issn}{0925-8388},
  \urlprefix\url{https://www.sciencedirect.com/science/article/pii/S092583889602467X}.

\bibitem[{\citenamefont{Mitsuda et~al.}(2012)\citenamefont{Mitsuda, Hamano,
  Araoka, Yayama, and Wada}}]{Mitsuda2012}
\bibinfo{author}{\bibfnamefont{A.}~\bibnamefont{Mitsuda}},
  \bibinfo{author}{\bibfnamefont{S.}~\bibnamefont{Hamano}},
  \bibinfo{author}{\bibfnamefont{N.}~\bibnamefont{Araoka}},
  \bibinfo{author}{\bibfnamefont{H.}~\bibnamefont{Yayama}}, \bibnamefont{and}
  \bibinfo{author}{\bibfnamefont{H.}~\bibnamefont{Wada}},
  \bibinfo{journal}{Journal of the Physical Society of Japan}
  \textbf{\bibinfo{volume}{81}}, \bibinfo{pages}{023709}
  (\bibinfo{year}{2012}), \eprint{https://doi.org/10.1143/JPSJ.81.023709},
  \urlprefix\url{https://doi.org/10.1143/JPSJ.81.023709}.

\bibitem[{\citenamefont{Bauminger et~al.}(1973)\citenamefont{Bauminger,
  Froindlich, Nowik, Ofer, Felner, and Mayer}}]{Bauminger1973}
\bibinfo{author}{\bibfnamefont{E.~R.} \bibnamefont{Bauminger}},
  \bibinfo{author}{\bibfnamefont{D.}~\bibnamefont{Froindlich}},
  \bibinfo{author}{\bibfnamefont{I.}~\bibnamefont{Nowik}},
  \bibinfo{author}{\bibfnamefont{S.}~\bibnamefont{Ofer}},
  \bibinfo{author}{\bibfnamefont{I.}~\bibnamefont{Felner}}, \bibnamefont{and}
  \bibinfo{author}{\bibfnamefont{I.}~\bibnamefont{Mayer}},
  \bibinfo{journal}{Phys. Rev. Lett.} \textbf{\bibinfo{volume}{30}},
  \bibinfo{pages}{1053} (\bibinfo{year}{1973}),
  \urlprefix\url{https://link.aps.org/doi/10.1103/PhysRevLett.30.1053}.

\bibitem[{\citenamefont{Wada et~al.}(1999)\citenamefont{Wada, Hundley,
  Movshovich, and Thompson}}]{Wada1999}
\bibinfo{author}{\bibfnamefont{H.}~\bibnamefont{Wada}},
  \bibinfo{author}{\bibfnamefont{M.~F.} \bibnamefont{Hundley}},
  \bibinfo{author}{\bibfnamefont{R.}~\bibnamefont{Movshovich}},
  \bibnamefont{and} \bibinfo{author}{\bibfnamefont{J.~D.}
  \bibnamefont{Thompson}}, \bibinfo{journal}{Phys. Rev. B}
  \textbf{\bibinfo{volume}{59}}, \bibinfo{pages}{1141} (\bibinfo{year}{1999}),
  \urlprefix\url{https://link.aps.org/doi/10.1103/PhysRevB.59.1141}.

\bibitem[{\citenamefont{Mitsuda et~al.}(2007)\citenamefont{Mitsuda, Ikeda,
  Ietaka, Fukuda, and Isikawa}}]{Mitsuda2007}
\bibinfo{author}{\bibfnamefont{A.}~\bibnamefont{Mitsuda}},
  \bibinfo{author}{\bibfnamefont{Y.}~\bibnamefont{Ikeda}},
  \bibinfo{author}{\bibfnamefont{N.}~\bibnamefont{Ietaka}},
  \bibinfo{author}{\bibfnamefont{S.}~\bibnamefont{Fukuda}}, \bibnamefont{and}
  \bibinfo{author}{\bibfnamefont{Y.}~\bibnamefont{Isikawa}},
  \bibinfo{journal}{Journal of Magnetism and Magnetic Materials}
  \textbf{\bibinfo{volume}{310}}, \bibinfo{pages}{319} (\bibinfo{year}{2007}),
  ISSN \bibinfo{issn}{0304-8853}, \bibinfo{note}{proceedings of the 17th
  International Conference on Magnetism},
  \urlprefix\url{https://www.sciencedirect.com/science/article/pii/S0304885306011632}.

\bibitem[{\citenamefont{Shannon}(1976)}]{Shannon1976}
\bibinfo{author}{\bibfnamefont{R.~D.} \bibnamefont{Shannon}},
  \bibinfo{journal}{Acta Crystallographica Section A}
  \textbf{\bibinfo{volume}{32}}, \bibinfo{pages}{751} (\bibinfo{year}{1976}),
  \urlprefix\url{https://doi.org/10.1107/S0567739476001551}.

\bibitem[{\citenamefont{{\=O}nuki et~al.}(2020)\citenamefont{{\=O}nuki, Hedo,
  and Honda}}]{Onuki2020}
\bibinfo{author}{\bibfnamefont{Y.}~\bibnamefont{{\=O}nuki}},
  \bibinfo{author}{\bibfnamefont{M.}~\bibnamefont{Hedo}}, \bibnamefont{and}
  \bibinfo{author}{\bibfnamefont{F.}~\bibnamefont{Honda}},
  \bibinfo{journal}{journal of the physical society of japan}
  \textbf{\bibinfo{volume}{89}}, \bibinfo{pages}{102001}
  (\bibinfo{year}{2020}).

\bibitem[{\citenamefont{Honda et~al.}(2017)\citenamefont{Honda, Okauchi,
  Nakamura, Aoki, Akamine, Ashitomi, Hedo, Nakama, and {\=O}nuki}}]{Honda2017}
\bibinfo{author}{\bibfnamefont{F.}~\bibnamefont{Honda}},
  \bibinfo{author}{\bibfnamefont{K.}~\bibnamefont{Okauchi}},
  \bibinfo{author}{\bibfnamefont{A.}~\bibnamefont{Nakamura}},
  \bibinfo{author}{\bibfnamefont{D.}~\bibnamefont{Aoki}},
  \bibinfo{author}{\bibfnamefont{H.}~\bibnamefont{Akamine}},
  \bibinfo{author}{\bibfnamefont{Y.}~\bibnamefont{Ashitomi}},
  \bibinfo{author}{\bibfnamefont{M.}~\bibnamefont{Hedo}},
  \bibinfo{author}{\bibfnamefont{T.}~\bibnamefont{Nakama}}, \bibnamefont{and}
  \bibinfo{author}{\bibfnamefont{Y.}~\bibnamefont{{\=O}nuki}}, in
  \emph{\bibinfo{booktitle}{Journal of Physics: Conference Series}}
  (\bibinfo{organization}{IOP Publishing}, \bibinfo{year}{2017}), vol.
  \bibinfo{volume}{807}, p. \bibinfo{pages}{022004}.

\bibitem[{\citenamefont{Venturini et~al.}(1985)\citenamefont{Venturini,
  Méot-Meyer, Malaman, and Roques}}]{Venturini1985}
\bibinfo{author}{\bibfnamefont{G.}~\bibnamefont{Venturini}},
  \bibinfo{author}{\bibfnamefont{M.}~\bibnamefont{Méot-Meyer}},
  \bibinfo{author}{\bibfnamefont{B.}~\bibnamefont{Malaman}}, \bibnamefont{and}
  \bibinfo{author}{\bibfnamefont{B.}~\bibnamefont{Roques}},
  \bibinfo{journal}{Journal of the Less Common Metals}
  \textbf{\bibinfo{volume}{113}}, \bibinfo{pages}{197} (\bibinfo{year}{1985}),
  ISSN \bibinfo{issn}{0022-5088},
  \urlprefix\url{https://www.sciencedirect.com/science/article/pii/0022508885902772}.

\bibitem[{\citenamefont{Maurya et~al.}(2014)\citenamefont{Maurya, Bonville,
  Thamizhavel, and Dhar}}]{Maurya2014}
\bibinfo{author}{\bibfnamefont{A.}~\bibnamefont{Maurya}},
  \bibinfo{author}{\bibfnamefont{P.}~\bibnamefont{Bonville}},
  \bibinfo{author}{\bibfnamefont{A.}~\bibnamefont{Thamizhavel}},
  \bibnamefont{and} \bibinfo{author}{\bibfnamefont{S.~K.} \bibnamefont{Dhar}},
  \bibinfo{journal}{Journal of Physics: Condensed Matter}
  \textbf{\bibinfo{volume}{26}}, \bibinfo{pages}{216001}
  (\bibinfo{year}{2014}),
  \urlprefix\url{https://doi.org/10.1088/0953-8984/26/21/216001}.

\bibitem[{\citenamefont{Maurya et~al.}(2016)\citenamefont{Maurya, Bonville,
  Kulkarni, Thamizhavel, and Dhar}}]{Maurya2016}
\bibinfo{author}{\bibfnamefont{A.}~\bibnamefont{Maurya}},
  \bibinfo{author}{\bibfnamefont{P.}~\bibnamefont{Bonville}},
  \bibinfo{author}{\bibfnamefont{R.}~\bibnamefont{Kulkarni}},
  \bibinfo{author}{\bibfnamefont{A.}~\bibnamefont{Thamizhavel}},
  \bibnamefont{and} \bibinfo{author}{\bibfnamefont{S.}~\bibnamefont{Dhar}},
  \bibinfo{journal}{Journal of Magnetism and Magnetic Materials}
  \textbf{\bibinfo{volume}{401}}, \bibinfo{pages}{823} (\bibinfo{year}{2016}),
  ISSN \bibinfo{issn}{0304-8853},
  \urlprefix\url{https://www.sciencedirect.com/science/article/pii/S0304885315307654}.

\bibitem[{\citenamefont{Fabr\`eges et~al.}(2016)\citenamefont{Fabr\`eges,
  Gukasov, Bonville, Maurya, Thamizhavel, and Dhar}}]{Fabreges2016}
\bibinfo{author}{\bibfnamefont{X.}~\bibnamefont{Fabr\`eges}},
  \bibinfo{author}{\bibfnamefont{A.}~\bibnamefont{Gukasov}},
  \bibinfo{author}{\bibfnamefont{P.}~\bibnamefont{Bonville}},
  \bibinfo{author}{\bibfnamefont{A.}~\bibnamefont{Maurya}},
  \bibinfo{author}{\bibfnamefont{A.}~\bibnamefont{Thamizhavel}},
  \bibnamefont{and} \bibinfo{author}{\bibfnamefont{S.~K.} \bibnamefont{Dhar}},
  \bibinfo{journal}{Phys. Rev. B} \textbf{\bibinfo{volume}{93}},
  \bibinfo{pages}{214414} (\bibinfo{year}{2016}),
  \urlprefix\url{https://link.aps.org/doi/10.1103/PhysRevB.93.214414}.

\bibitem[{\citenamefont{Bednarchuk et~al.}(2015)\citenamefont{Bednarchuk,
  Gagor, and Kaczorowski}}]{Bednarchuk_JAC_2015}
\bibinfo{author}{\bibfnamefont{O.}~\bibnamefont{Bednarchuk}},
  \bibinfo{author}{\bibfnamefont{A.}~\bibnamefont{Gagor}}, \bibnamefont{and}
  \bibinfo{author}{\bibfnamefont{D.}~\bibnamefont{Kaczorowski}},
  \bibinfo{journal}{Journal of Alloys and Compounds}
  \textbf{\bibinfo{volume}{622}}, \bibinfo{pages}{432} (\bibinfo{year}{2015}).

\bibitem[{\citenamefont{Kakihana et~al.}(2017)\citenamefont{Kakihana, Akamine,
  Tomori, Nishimura, Teruya, Nakamura, Honda, Aoki, Nakashima, Amako
  et~al.}}]{Kakihana2017}
\bibinfo{author}{\bibfnamefont{M.}~\bibnamefont{Kakihana}},
  \bibinfo{author}{\bibfnamefont{H.}~\bibnamefont{Akamine}},
  \bibinfo{author}{\bibfnamefont{K.}~\bibnamefont{Tomori}},
  \bibinfo{author}{\bibfnamefont{K.}~\bibnamefont{Nishimura}},
  \bibinfo{author}{\bibfnamefont{A.}~\bibnamefont{Teruya}},
  \bibinfo{author}{\bibfnamefont{A.}~\bibnamefont{Nakamura}},
  \bibinfo{author}{\bibfnamefont{F.}~\bibnamefont{Honda}},
  \bibinfo{author}{\bibfnamefont{D.}~\bibnamefont{Aoki}},
  \bibinfo{author}{\bibfnamefont{M.}~\bibnamefont{Nakashima}},
  \bibinfo{author}{\bibfnamefont{Y.}~\bibnamefont{Amako}},
  \bibnamefont{et~al.}, \bibinfo{journal}{Journal of Alloys and Compounds}
  \textbf{\bibinfo{volume}{694}}, \bibinfo{pages}{439} (\bibinfo{year}{2017}),
  ISSN \bibinfo{issn}{0925-8388},
  \urlprefix\url{https://www.sciencedirect.com/science/article/pii/S092583881633050X}.

\bibitem[{\citenamefont{Bauer et~al.}(2022)\citenamefont{Bauer, Senyshyn,
  Bozhanova, Simeth, Franz, Gottlieb-Sch\"onmeyer, Meven, Schrader, and
  Pfleiderer}}]{Bauer2022}
\bibinfo{author}{\bibfnamefont{A.}~\bibnamefont{Bauer}},
  \bibinfo{author}{\bibfnamefont{A.}~\bibnamefont{Senyshyn}},
  \bibinfo{author}{\bibfnamefont{R.}~\bibnamefont{Bozhanova}},
  \bibinfo{author}{\bibfnamefont{W.}~\bibnamefont{Simeth}},
  \bibinfo{author}{\bibfnamefont{C.}~\bibnamefont{Franz}},
  \bibinfo{author}{\bibfnamefont{S.}~\bibnamefont{Gottlieb-Sch\"onmeyer}},
  \bibinfo{author}{\bibfnamefont{M.}~\bibnamefont{Meven}},
  \bibinfo{author}{\bibfnamefont{T.~E.} \bibnamefont{Schrader}},
  \bibnamefont{and}
  \bibinfo{author}{\bibfnamefont{C.}~\bibnamefont{Pfleiderer}},
  \bibinfo{journal}{Phys. Rev. Materials} \textbf{\bibinfo{volume}{6}},
  \bibinfo{pages}{034406} (\bibinfo{year}{2022}),
  \urlprefix\url{https://link.aps.org/doi/10.1103/PhysRevMaterials.6.034406}.

\bibitem[{\citenamefont{Matsumura et~al.}(2022)\citenamefont{Matsumura,
  Tsukagoshi, Ueda, Higa, Nakao, Kaneko, Kakihana, Hedo, Nakama, and
  {\=O}nuki}}]{Matsumura2022}
\bibinfo{author}{\bibfnamefont{T.}~\bibnamefont{Matsumura}},
  \bibinfo{author}{\bibfnamefont{M.}~\bibnamefont{Tsukagoshi}},
  \bibinfo{author}{\bibfnamefont{Y.}~\bibnamefont{Ueda}},
  \bibinfo{author}{\bibfnamefont{N.}~\bibnamefont{Higa}},
  \bibinfo{author}{\bibfnamefont{A.}~\bibnamefont{Nakao}},
  \bibinfo{author}{\bibfnamefont{K.}~\bibnamefont{Kaneko}},
  \bibinfo{author}{\bibfnamefont{M.}~\bibnamefont{Kakihana}},
  \bibinfo{author}{\bibfnamefont{M.}~\bibnamefont{Hedo}},
  \bibinfo{author}{\bibfnamefont{T.}~\bibnamefont{Nakama}}, \bibnamefont{and}
  \bibinfo{author}{\bibfnamefont{Y.}~\bibnamefont{{\=O}nuki}},
  \bibinfo{journal}{Journal of the Physical Society of Japan}
  \textbf{\bibinfo{volume}{91}}, \bibinfo{pages}{073703}
  (\bibinfo{year}{2022}), \eprint{https://doi.org/10.7566/JPSJ.91.073703},
  \urlprefix\url{https://doi.org/10.7566/JPSJ.91.073703}.

\bibitem[{\citenamefont{Nakashima et~al.}(2017)\citenamefont{Nakashima, Amako,
  Matsubayashi, Uwatoko, Nada, Sugiyama, Hagiwara, Haga, Takeuchi, Nakamura
  et~al.}}]{Nakashima_JPSJ_2017}
\bibinfo{author}{\bibfnamefont{M.}~\bibnamefont{Nakashima}},
  \bibinfo{author}{\bibfnamefont{Y.}~\bibnamefont{Amako}},
  \bibinfo{author}{\bibfnamefont{K.}~\bibnamefont{Matsubayashi}},
  \bibinfo{author}{\bibfnamefont{Y.}~\bibnamefont{Uwatoko}},
  \bibinfo{author}{\bibfnamefont{M.}~\bibnamefont{Nada}},
  \bibinfo{author}{\bibfnamefont{K.}~\bibnamefont{Sugiyama}},
  \bibinfo{author}{\bibfnamefont{M.}~\bibnamefont{Hagiwara}},
  \bibinfo{author}{\bibfnamefont{Y.}~\bibnamefont{Haga}},
  \bibinfo{author}{\bibfnamefont{T.}~\bibnamefont{Takeuchi}},
  \bibinfo{author}{\bibfnamefont{A.}~\bibnamefont{Nakamura}},
  \bibnamefont{et~al.}, \bibinfo{journal}{Journal of the Physical Society of
  Japan} \textbf{\bibinfo{volume}{86}}, \bibinfo{pages}{034708}
  (\bibinfo{year}{2017}), \eprint{https://doi.org/10.7566/JPSJ.86.034708},
  \urlprefix\url{https://doi.org/10.7566/JPSJ.86.034708}.

\bibitem[{\citenamefont{Nakamura et~al.}(2015)\citenamefont{Nakamura, Okazaki,
  Nakashima, Amako, Matsubayashi, Uwatoko, Kayama, Kagayama, Shimizu, Uejo
  et~al.}}]{Nakamura2015}
\bibinfo{author}{\bibfnamefont{A.}~\bibnamefont{Nakamura}},
  \bibinfo{author}{\bibfnamefont{T.}~\bibnamefont{Okazaki}},
  \bibinfo{author}{\bibfnamefont{M.}~\bibnamefont{Nakashima}},
  \bibinfo{author}{\bibfnamefont{Y.}~\bibnamefont{Amako}},
  \bibinfo{author}{\bibfnamefont{K.}~\bibnamefont{Matsubayashi}},
  \bibinfo{author}{\bibfnamefont{Y.}~\bibnamefont{Uwatoko}},
  \bibinfo{author}{\bibfnamefont{S.}~\bibnamefont{Kayama}},
  \bibinfo{author}{\bibfnamefont{T.}~\bibnamefont{Kagayama}},
  \bibinfo{author}{\bibfnamefont{K.}~\bibnamefont{Shimizu}},
  \bibinfo{author}{\bibfnamefont{T.}~\bibnamefont{Uejo}}, \bibnamefont{et~al.},
  \bibinfo{journal}{Journal of the Physical Society of Japan}
  \textbf{\bibinfo{volume}{84}}, \bibinfo{pages}{053701}
  (\bibinfo{year}{2015}), \eprint{https://doi.org/10.7566/JPSJ.84.053701},
  \urlprefix\url{https://doi.org/10.7566/JPSJ.84.053701}.

\bibitem[{\citenamefont{Kimura et~al.}(2005)\citenamefont{Kimura, Ito, Saitoh,
  Umeda, Aoki, and Terashima}}]{Kimura2005}
\bibinfo{author}{\bibfnamefont{N.}~\bibnamefont{Kimura}},
  \bibinfo{author}{\bibfnamefont{K.}~\bibnamefont{Ito}},
  \bibinfo{author}{\bibfnamefont{K.}~\bibnamefont{Saitoh}},
  \bibinfo{author}{\bibfnamefont{Y.}~\bibnamefont{Umeda}},
  \bibinfo{author}{\bibfnamefont{H.}~\bibnamefont{Aoki}}, \bibnamefont{and}
  \bibinfo{author}{\bibfnamefont{T.}~\bibnamefont{Terashima}},
  \bibinfo{journal}{Phys. Rev. Lett.} \textbf{\bibinfo{volume}{95}},
  \bibinfo{pages}{247004} (\bibinfo{year}{2005}),
  \urlprefix\url{https://link.aps.org/doi/10.1103/PhysRevLett.95.247004}.

\bibitem[{\citenamefont{Sugitani et~al.}(2006)\citenamefont{Sugitani, Okuda,
  Shishido, Yamada, Thamizhavel, Yamamoto, D.~Matsuda, Haga, Takeuchi, Settai
  et~al.}}]{Sugitani2006}
\bibinfo{author}{\bibfnamefont{I.}~\bibnamefont{Sugitani}},
  \bibinfo{author}{\bibfnamefont{Y.}~\bibnamefont{Okuda}},
  \bibinfo{author}{\bibfnamefont{H.}~\bibnamefont{Shishido}},
  \bibinfo{author}{\bibfnamefont{T.}~\bibnamefont{Yamada}},
  \bibinfo{author}{\bibfnamefont{A.}~\bibnamefont{Thamizhavel}},
  \bibinfo{author}{\bibfnamefont{E.}~\bibnamefont{Yamamoto}},
  \bibinfo{author}{\bibfnamefont{T.}~\bibnamefont{D.~Matsuda}},
  \bibinfo{author}{\bibfnamefont{Y.}~\bibnamefont{Haga}},
  \bibinfo{author}{\bibfnamefont{T.}~\bibnamefont{Takeuchi}},
  \bibinfo{author}{\bibfnamefont{R.}~\bibnamefont{Settai}},
  \bibnamefont{et~al.}, \bibinfo{journal}{Journal of the Physical Society of
  Japan} \textbf{\bibinfo{volume}{75}}, \bibinfo{pages}{043703}
  (\bibinfo{year}{2006}), \eprint{https://doi.org/10.1143/JPSJ.75.043703},
  \urlprefix\url{https://doi.org/10.1143/JPSJ.75.043703}.

\bibitem[{\citenamefont{Settai et~al.}(2007)\citenamefont{Settai, Sugitani,
  Okuda, Thamizhavel, Nakashima, {\=O}nuki, and Harima}}]{Settai2007}
\bibinfo{author}{\bibfnamefont{R.}~\bibnamefont{Settai}},
  \bibinfo{author}{\bibfnamefont{I.}~\bibnamefont{Sugitani}},
  \bibinfo{author}{\bibfnamefont{Y.}~\bibnamefont{Okuda}},
  \bibinfo{author}{\bibfnamefont{A.}~\bibnamefont{Thamizhavel}},
  \bibinfo{author}{\bibfnamefont{M.}~\bibnamefont{Nakashima}},
  \bibinfo{author}{\bibfnamefont{Y.}~\bibnamefont{{\=O}nuki}},
  \bibnamefont{and} \bibinfo{author}{\bibfnamefont{H.}~\bibnamefont{Harima}},
  \bibinfo{journal}{Journal of magnetism and magnetic materials}
  \textbf{\bibinfo{volume}{310}}, \bibinfo{pages}{844} (\bibinfo{year}{2007}).

\bibitem[{\citenamefont{Honda et~al.}(2010)\citenamefont{Honda, Bonalde,
  Yoshiuchi, Hirose, Nakamura, Shimizu, Settai, and {\=O}nuki}}]{Honda2010}
\bibinfo{author}{\bibfnamefont{F.}~\bibnamefont{Honda}},
  \bibinfo{author}{\bibfnamefont{I.}~\bibnamefont{Bonalde}},
  \bibinfo{author}{\bibfnamefont{S.}~\bibnamefont{Yoshiuchi}},
  \bibinfo{author}{\bibfnamefont{Y.}~\bibnamefont{Hirose}},
  \bibinfo{author}{\bibfnamefont{T.}~\bibnamefont{Nakamura}},
  \bibinfo{author}{\bibfnamefont{K.}~\bibnamefont{Shimizu}},
  \bibinfo{author}{\bibfnamefont{R.}~\bibnamefont{Settai}}, \bibnamefont{and}
  \bibinfo{author}{\bibfnamefont{Y.}~\bibnamefont{{\=O}nuki}},
  \bibinfo{journal}{Physica C: Superconductivity and its applications}
  \textbf{\bibinfo{volume}{470}}, \bibinfo{pages}{S543} (\bibinfo{year}{2010}).

\bibitem[{\citenamefont{Takimoto}(2008)}]{Takimoto_JPSJ_2008}
\bibinfo{author}{\bibfnamefont{T.}~\bibnamefont{Takimoto}},
  \bibinfo{journal}{Journal of the Physical Society of Japan}
  \textbf{\bibinfo{volume}{77}}, \bibinfo{pages}{113706}
  (\bibinfo{year}{2008}), \eprint{https://doi.org/10.1143/JPSJ.77.113706},
  \urlprefix\url{https://doi.org/10.1143/JPSJ.77.113706}.

\bibitem[{\citenamefont{Takimoto and Thalmeier}(2009)}]{Takimoto_JPSJ_2009}
\bibinfo{author}{\bibfnamefont{T.}~\bibnamefont{Takimoto}} \bibnamefont{and}
  \bibinfo{author}{\bibfnamefont{P.}~\bibnamefont{Thalmeier}},
  \bibinfo{journal}{Journal of the Physical Society of Japan}
  \textbf{\bibinfo{volume}{78}}, \bibinfo{pages}{103703}
  (\bibinfo{year}{2009}).

\bibitem[{\citenamefont{Bednarchuk and
  Kaczorowski}(2015{\natexlab{a}})}]{Bednarchuk_JAC_2_2015}
\bibinfo{author}{\bibfnamefont{O.}~\bibnamefont{Bednarchuk}} \bibnamefont{and}
  \bibinfo{author}{\bibfnamefont{D.}~\bibnamefont{Kaczorowski}},
  \bibinfo{journal}{Journal of Alloys and Compounds}
  \textbf{\bibinfo{volume}{646}}, \bibinfo{pages}{291}
  (\bibinfo{year}{2015}{\natexlab{a}}).

\bibitem[{\citenamefont{Bednarchuk and
  Kaczorowski}(2015{\natexlab{b}})}]{Bednarchuk_APPA_2015}
\bibinfo{author}{\bibfnamefont{O.}~\bibnamefont{Bednarchuk}} \bibnamefont{and}
  \bibinfo{author}{\bibfnamefont{D.}~\bibnamefont{Kaczorowski}},
  \bibinfo{journal}{Acta Physica Polonica A} \textbf{\bibinfo{volume}{127}},
  \bibinfo{pages}{418} (\bibinfo{year}{2015}{\natexlab{b}}).

\bibitem[{\citenamefont{Uchima et~al.}(2014{\natexlab{a}})\citenamefont{Uchima,
  Arakaki, Hirakawa, Hiranaka, Uejo, Teruya, Nakamura, Takeda, Takaesu, Hedo
  et~al.}}]{Uchima2014}
\bibinfo{author}{\bibfnamefont{K.}~\bibnamefont{Uchima}},
  \bibinfo{author}{\bibfnamefont{N.}~\bibnamefont{Arakaki}},
  \bibinfo{author}{\bibfnamefont{S.}~\bibnamefont{Hirakawa}},
  \bibinfo{author}{\bibfnamefont{Y.}~\bibnamefont{Hiranaka}},
  \bibinfo{author}{\bibfnamefont{T.}~\bibnamefont{Uejo}},
  \bibinfo{author}{\bibfnamefont{A.}~\bibnamefont{Teruya}},
  \bibinfo{author}{\bibfnamefont{A.}~\bibnamefont{Nakamura}},
  \bibinfo{author}{\bibfnamefont{M.}~\bibnamefont{Takeda}},
  \bibinfo{author}{\bibfnamefont{Y.}~\bibnamefont{Takaesu}},
  \bibinfo{author}{\bibfnamefont{M.}~\bibnamefont{Hedo}}, \bibnamefont{et~al.},
  \emph{\bibinfo{title}{Pressure Effect on Transport Properties of EuNiGe3}}
  (\bibinfo{year}{2014}{\natexlab{a}}),
  \eprint{https://journals.jps.jp/doi/pdf/10.7566/JPSCP.1.012015},
  \urlprefix\url{https://journals.jps.jp/doi/abs/10.7566/JPSCP.1.012015}.

\bibitem[{\citenamefont{Muthu et~al.}(2019)\citenamefont{Muthu, Braithwaite,
  Salce, Arumugam, Govindaraj, Saravanan, Kanagaraj, Sarkar, and
  Peter}}]{Muthu2019}
\bibinfo{author}{\bibfnamefont{S.~E.} \bibnamefont{Muthu}},
  \bibinfo{author}{\bibfnamefont{D.}~\bibnamefont{Braithwaite}},
  \bibinfo{author}{\bibfnamefont{B.}~\bibnamefont{Salce}},
  \bibinfo{author}{\bibfnamefont{S.}~\bibnamefont{Arumugam}},
  \bibinfo{author}{\bibfnamefont{L.}~\bibnamefont{Govindaraj}},
  \bibinfo{author}{\bibfnamefont{C.}~\bibnamefont{Saravanan}},
  \bibinfo{author}{\bibfnamefont{M.}~\bibnamefont{Kanagaraj}},
  \bibinfo{author}{\bibfnamefont{S.}~\bibnamefont{Sarkar}}, \bibnamefont{and}
  \bibinfo{author}{\bibfnamefont{S.~C.} \bibnamefont{Peter}},
  \bibinfo{journal}{Journal of the Physical Society of Japan}
  \textbf{\bibinfo{volume}{88}}, \bibinfo{pages}{074702}
  (\bibinfo{year}{2019}), \eprint{https://doi.org/10.7566/JPSJ.88.074702},
  \urlprefix\url{https://doi.org/10.7566/JPSJ.88.074702}.

\bibitem[{\citenamefont{Rueff et~al.}(2015)\citenamefont{Rueff, Ablett,
  C{\'{e}}olin, Prieur, Moreno, Bal{\'{e}}dent, Lassalle-Kaiser, Rault, Simon,
  and Shukla}}]{Rueff_2014}
\bibinfo{author}{\bibfnamefont{J.-P.} \bibnamefont{Rueff}},
  \bibinfo{author}{\bibfnamefont{J.~M.} \bibnamefont{Ablett}},
  \bibinfo{author}{\bibfnamefont{D.}~\bibnamefont{C{\'{e}}olin}},
  \bibinfo{author}{\bibfnamefont{D.}~\bibnamefont{Prieur}},
  \bibinfo{author}{\bibfnamefont{T.}~\bibnamefont{Moreno}},
  \bibinfo{author}{\bibfnamefont{V.}~\bibnamefont{Bal{\'{e}}dent}},
  \bibinfo{author}{\bibfnamefont{B.}~\bibnamefont{Lassalle-Kaiser}},
  \bibinfo{author}{\bibfnamefont{J.~E.} \bibnamefont{Rault}},
  \bibinfo{author}{\bibfnamefont{M.}~\bibnamefont{Simon}}, \bibnamefont{and}
  \bibinfo{author}{\bibfnamefont{A.}~\bibnamefont{Shukla}},
  \bibinfo{journal}{Journal of Synchrotron Radiation}
  \textbf{\bibinfo{volume}{22}}, \bibinfo{pages}{175} (\bibinfo{year}{2015}),
  \urlprefix\url{https://doi.org/10.1107/S160057751402102X}.

\bibitem[{\citenamefont{Ablett et~al.}(2019)\citenamefont{Ablett, Prieur,
  C{\'{e}}olin, Lassalle-Kaiser, Lebert, Sauvage, Moreno, Bac, Bal{\'{e}}dent,
  Ovono et~al.}}]{Ablett_2019}
\bibinfo{author}{\bibfnamefont{J.~M.} \bibnamefont{Ablett}},
  \bibinfo{author}{\bibfnamefont{D.}~\bibnamefont{Prieur}},
  \bibinfo{author}{\bibfnamefont{D.}~\bibnamefont{C{\'{e}}olin}},
  \bibinfo{author}{\bibfnamefont{B.}~\bibnamefont{Lassalle-Kaiser}},
  \bibinfo{author}{\bibfnamefont{B.}~\bibnamefont{Lebert}},
  \bibinfo{author}{\bibfnamefont{M.}~\bibnamefont{Sauvage}},
  \bibinfo{author}{\bibfnamefont{T.}~\bibnamefont{Moreno}},
  \bibinfo{author}{\bibfnamefont{S.}~\bibnamefont{Bac}},
  \bibinfo{author}{\bibfnamefont{V.}~\bibnamefont{Bal{\'{e}}dent}},
  \bibinfo{author}{\bibfnamefont{A.}~\bibnamefont{Ovono}},
  \bibnamefont{et~al.}, \bibinfo{journal}{Journal of Synchrotron Radiation}
  \textbf{\bibinfo{volume}{26}}, \bibinfo{pages}{263} (\bibinfo{year}{2019}),
  \urlprefix\url{https://doi.org/10.1107/S160057751801559X}.

\bibitem[{\citenamefont{Ablett et~al.}(2021)\citenamefont{Ablett, Shieh,
  Bal\'edent, Woicik, Cockayne, and Shirley}}]{James_2021}
\bibinfo{author}{\bibfnamefont{J.~M.} \bibnamefont{Ablett}},
  \bibinfo{author}{\bibfnamefont{S.~R.} \bibnamefont{Shieh}},
  \bibinfo{author}{\bibfnamefont{V.}~\bibnamefont{Bal\'edent}},
  \bibinfo{author}{\bibfnamefont{J.~C.} \bibnamefont{Woicik}},
  \bibinfo{author}{\bibfnamefont{E.}~\bibnamefont{Cockayne}}, \bibnamefont{and}
  \bibinfo{author}{\bibfnamefont{E.~L.} \bibnamefont{Shirley}},
  \bibinfo{journal}{Phys. Rev. B} \textbf{\bibinfo{volume}{104}},
  \bibinfo{pages}{054119} (\bibinfo{year}{2021}),
  \urlprefix\url{https://link.aps.org/doi/10.1103/PhysRevB.104.054119}.

\bibitem[{\citenamefont{Mao et~al.}(1986)\citenamefont{Mao, Xu, and
  Bell}}]{Mao_1986}
\bibinfo{author}{\bibfnamefont{H.~K.} \bibnamefont{Mao}},
  \bibinfo{author}{\bibfnamefont{J.}~\bibnamefont{Xu}}, \bibnamefont{and}
  \bibinfo{author}{\bibfnamefont{P.~M.} \bibnamefont{Bell}},
  \bibinfo{journal}{Journal of Geophysical Research: Solid Earth}
  \textbf{\bibinfo{volume}{91}}, \bibinfo{pages}{4673} (\bibinfo{year}{1986}),
  \urlprefix\url{https://agupubs.onlinelibrary.wiley.com/doi/abs/10.1029/JB091iB05p04673}.

\bibitem[{\citenamefont{Utsumi et~al.}(2021)\citenamefont{Utsumi,
  Batisti{\'{c}}, Bal{\'{e}}dent, Shieh, Dhami, Bednarchuk, Kaczorowski,
  Ablett, and Rueff}}]{Utsumi_ES_2021}
\bibinfo{author}{\bibfnamefont{Y.}~\bibnamefont{Utsumi}},
  \bibinfo{author}{\bibfnamefont{I.}~\bibnamefont{Batisti{\'{c}}}},
  \bibinfo{author}{\bibfnamefont{V.}~\bibnamefont{Bal{\'{e}}dent}},
  \bibinfo{author}{\bibfnamefont{S.~R.} \bibnamefont{Shieh}},
  \bibinfo{author}{\bibfnamefont{N.~S.} \bibnamefont{Dhami}},
  \bibinfo{author}{\bibfnamefont{O.}~\bibnamefont{Bednarchuk}},
  \bibinfo{author}{\bibfnamefont{D.}~\bibnamefont{Kaczorowski}},
  \bibinfo{author}{\bibfnamefont{J.~M.} \bibnamefont{Ablett}},
  \bibnamefont{and} \bibinfo{author}{\bibfnamefont{J.~P.} \bibnamefont{Rueff}},
  \bibinfo{journal}{Electronic Structure} \textbf{\bibinfo{volume}{3}},
  \bibinfo{pages}{034002} (\bibinfo{year}{2021}),
  \urlprefix\url{https://doi.org/10.1088/2516-1075/ac0c27}.

\bibitem[{\citenamefont{{\=O}nuki et~al.}(2017)\citenamefont{{\=O}nuki,
  Nakamura, Honda, Aoki, Tekeuchi, Nakashima, Amako, Harima, Matsubayashi,
  Uwatoko et~al.}}]{Onuki_PM_2017}
\bibinfo{author}{\bibfnamefont{Y.}~\bibnamefont{{\=O}nuki}},
  \bibinfo{author}{\bibfnamefont{A.}~\bibnamefont{Nakamura}},
  \bibinfo{author}{\bibfnamefont{F.}~\bibnamefont{Honda}},
  \bibinfo{author}{\bibfnamefont{D.}~\bibnamefont{Aoki}},
  \bibinfo{author}{\bibfnamefont{T.}~\bibnamefont{Tekeuchi}},
  \bibinfo{author}{\bibfnamefont{M.}~\bibnamefont{Nakashima}},
  \bibinfo{author}{\bibfnamefont{Y.}~\bibnamefont{Amako}},
  \bibinfo{author}{\bibfnamefont{H.}~\bibnamefont{Harima}},
  \bibinfo{author}{\bibfnamefont{K.}~\bibnamefont{Matsubayashi}},
  \bibinfo{author}{\bibfnamefont{Y.}~\bibnamefont{Uwatoko}},
  \bibnamefont{et~al.}, \bibinfo{journal}{Philosophical Magazine}
  \textbf{\bibinfo{volume}{97}}, \bibinfo{pages}{3399} (\bibinfo{year}{2017}),
  \eprint{https://doi.org/10.1080/14786435.2016.1218081},
  \urlprefix\url{https://doi.org/10.1080/14786435.2016.1218081}.

\bibitem[{\citenamefont{Hlil et~al.}(1996)\citenamefont{Hlil, Baudoing-Savois,
  Moraweck, and Renouprez}}]{Hlil1996}
\bibinfo{author}{\bibfnamefont{E.~K.} \bibnamefont{Hlil}},
  \bibinfo{author}{\bibfnamefont{R.}~\bibnamefont{Baudoing-Savois}},
  \bibinfo{author}{\bibfnamefont{B.}~\bibnamefont{Moraweck}}, \bibnamefont{and}
  \bibinfo{author}{\bibfnamefont{A.~J.} \bibnamefont{Renouprez}},
  \bibinfo{journal}{The Journal of Physical Chemistry}
  \textbf{\bibinfo{volume}{100}}, \bibinfo{pages}{3102} (\bibinfo{year}{1996}),
  \eprint{https://doi.org/10.1021/jp951440t},
  \urlprefix\url{https://doi.org/10.1021/jp951440t}.

\bibitem[{\citenamefont{Vank\'o et~al.}(2006)\citenamefont{Vank\'o, Rueff,
  Mattila, N\'emeth, and Shukla}}]{Vanko2006}
\bibinfo{author}{\bibfnamefont{G.}~\bibnamefont{Vank\'o}},
  \bibinfo{author}{\bibfnamefont{J.-P.} \bibnamefont{Rueff}},
  \bibinfo{author}{\bibfnamefont{A.}~\bibnamefont{Mattila}},
  \bibinfo{author}{\bibfnamefont{Z.}~\bibnamefont{N\'emeth}}, \bibnamefont{and}
  \bibinfo{author}{\bibfnamefont{A.}~\bibnamefont{Shukla}},
  \bibinfo{journal}{Phys. Rev. B} \textbf{\bibinfo{volume}{73}},
  \bibinfo{pages}{024424} (\bibinfo{year}{2006}),
  \urlprefix\url{https://link.aps.org/doi/10.1103/PhysRevB.73.024424}.

\bibitem[{\citenamefont{Kim et~al.}(1997)\citenamefont{Kim, Im, Oh, Kim, and
  Yo}}]{Kim1997}
\bibinfo{author}{\bibfnamefont{M.}~\bibnamefont{Kim}},
  \bibinfo{author}{\bibfnamefont{Y.}~\bibnamefont{Im}},
  \bibinfo{author}{\bibfnamefont{E.}~\bibnamefont{Oh}},
  \bibinfo{author}{\bibfnamefont{K.}~\bibnamefont{Kim}}, \bibnamefont{and}
  \bibinfo{author}{\bibfnamefont{C.}~\bibnamefont{Yo}},
  \bibinfo{journal}{Physica B: Condensed Matter}
  \textbf{\bibinfo{volume}{229}}, \bibinfo{pages}{338} (\bibinfo{year}{1997}),
  ISSN \bibinfo{issn}{0921-4526},
  \urlprefix\url{https://www.sciencedirect.com/science/article/pii/S0921452696008484}.

\bibitem[{\citenamefont{Rogalev et~al.}(2021)\citenamefont{Rogalev, Wilhelm,
  Ovchinnikova, Enikeev, Bakonin, Kozlovskaya, Oreshko, Aoki, and
  Dmitrienko}}]{Rogalev2021}
\bibinfo{author}{\bibfnamefont{A.}~\bibnamefont{Rogalev}},
  \bibinfo{author}{\bibfnamefont{F.}~\bibnamefont{Wilhelm}},
  \bibinfo{author}{\bibfnamefont{E.}~\bibnamefont{Ovchinnikova}},
  \bibinfo{author}{\bibfnamefont{A.}~\bibnamefont{Enikeev}},
  \bibinfo{author}{\bibfnamefont{R.}~\bibnamefont{Bakonin}},
  \bibinfo{author}{\bibfnamefont{K.}~\bibnamefont{Kozlovskaya}},
  \bibinfo{author}{\bibfnamefont{A.}~\bibnamefont{Oreshko}},
  \bibinfo{author}{\bibfnamefont{D.}~\bibnamefont{Aoki}}, \bibnamefont{and}
  \bibinfo{author}{\bibfnamefont{V.~E.} \bibnamefont{Dmitrienko}},
  \bibinfo{journal}{Crystals} \textbf{\bibinfo{volume}{11}}
  (\bibinfo{year}{2021}), ISSN \bibinfo{issn}{2073-4352},
  \urlprefix\url{https://www.mdpi.com/2073-4352/11/5/544}.

\bibitem[{\citenamefont{Finger et~al.}(1981)\citenamefont{Finger, Hazen, Zou,
  Mao, and Bell}}]{Finger1981}
\bibinfo{author}{\bibfnamefont{L.~W.} \bibnamefont{Finger}},
  \bibinfo{author}{\bibfnamefont{R.~M.} \bibnamefont{Hazen}},
  \bibinfo{author}{\bibfnamefont{G.}~\bibnamefont{Zou}},
  \bibinfo{author}{\bibfnamefont{H.~K.} \bibnamefont{Mao}}, \bibnamefont{and}
  \bibinfo{author}{\bibfnamefont{P.~M.} \bibnamefont{Bell}},
  \bibinfo{journal}{Applied Physics Letters} \textbf{\bibinfo{volume}{39}},
  \bibinfo{pages}{892} (\bibinfo{year}{1981}),
  \eprint{https://doi.org/10.1063/1.92597},
  \urlprefix\url{https://doi.org/10.1063/1.92597}.

\bibitem[{\citenamefont{Meng et~al.}(1993)\citenamefont{Meng, Weidner, and
  Fei}}]{Meng1993}
\bibinfo{author}{\bibfnamefont{Y.}~\bibnamefont{Meng}},
  \bibinfo{author}{\bibfnamefont{D.~J.} \bibnamefont{Weidner}},
  \bibnamefont{and} \bibinfo{author}{\bibfnamefont{Y.}~\bibnamefont{Fei}},
  \bibinfo{journal}{Geophysical Research Letters}
  \textbf{\bibinfo{volume}{20}}, \bibinfo{pages}{1147} (\bibinfo{year}{1993}),
  \urlprefix\url{https://agupubs.onlinelibrary.wiley.com
  /doi/abs/10.1029/93GL01400}.

\bibitem[{\citenamefont{Klotz et~al.}(2009)\citenamefont{Klotz, Chervin,
  Munsch, and Marchand}}]{Klotz_IOP_2009}
\bibinfo{author}{\bibfnamefont{S.}~\bibnamefont{Klotz}},
  \bibinfo{author}{\bibfnamefont{J.-C.} \bibnamefont{Chervin}},
  \bibinfo{author}{\bibfnamefont{P.}~\bibnamefont{Munsch}}, \bibnamefont{and}
  \bibinfo{author}{\bibfnamefont{G.~L.} \bibnamefont{Marchand}},
  \bibinfo{journal}{Journal of Physics D: Applied Physics}
  \textbf{\bibinfo{volume}{42}}, \bibinfo{pages}{075413}
  (\bibinfo{year}{2009}),
  \urlprefix\url{https://doi.org/10.1088/0022-3727/42/7/075413}.

\bibitem[{\citenamefont{Doebelin and Kleeberg}(2015)}]{Profex}
\bibinfo{author}{\bibfnamefont{N.}~\bibnamefont{Doebelin}} \bibnamefont{and}
  \bibinfo{author}{\bibfnamefont{R.}~\bibnamefont{Kleeberg}},
  \bibinfo{journal}{Journal of Applied Crystallography}
  \textbf{\bibinfo{volume}{48}}, \bibinfo{pages}{1573} (\bibinfo{year}{2015}),
  \urlprefix\url{https://doi.org/10.1107/S1600576715014685}.

\bibitem[{\citenamefont{Hoffmann and Zheng}(1985)}]{Hoffmann1985}
\bibinfo{author}{\bibfnamefont{R.}~\bibnamefont{Hoffmann}} \bibnamefont{and}
  \bibinfo{author}{\bibfnamefont{C.}~\bibnamefont{Zheng}},
  \bibinfo{journal}{The Journal of Physical Chemistry}
  \textbf{\bibinfo{volume}{89}}, \bibinfo{pages}{4175} (\bibinfo{year}{1985}),
  \eprint{https://doi.org/10.1021/j100266a007},
  \urlprefix\url{https://doi.org/10.1021/j100266a007}.

\bibitem[{\citenamefont{Johrendt et~al.}(1997)\citenamefont{Johrendt, Felser,
  Jepsen, Andersen, Mewis, and Rouxel}}]{Johrendt1997}
\bibinfo{author}{\bibfnamefont{D.}~\bibnamefont{Johrendt}},
  \bibinfo{author}{\bibfnamefont{C.}~\bibnamefont{Felser}},
  \bibinfo{author}{\bibfnamefont{O.}~\bibnamefont{Jepsen}},
  \bibinfo{author}{\bibfnamefont{O.~K.} \bibnamefont{Andersen}},
  \bibinfo{author}{\bibfnamefont{A.}~\bibnamefont{Mewis}}, \bibnamefont{and}
  \bibinfo{author}{\bibfnamefont{J.}~\bibnamefont{Rouxel}},
  \bibinfo{journal}{Journal of Solid State Chemistry}
  \textbf{\bibinfo{volume}{130}}, \bibinfo{pages}{254} (\bibinfo{year}{1997}),
  ISSN \bibinfo{issn}{0022-4596},
  \urlprefix\url{https://www.sciencedirect.com/science/article/pii/S002245969797300X}.

\bibitem[{\citenamefont{Huhnt et~al.}(1998)\citenamefont{Huhnt, Schlabitz,
  Wurth, Mewis, and Reehuis}}]{Huhnt1998}
\bibinfo{author}{\bibfnamefont{C.}~\bibnamefont{Huhnt}},
  \bibinfo{author}{\bibfnamefont{W.}~\bibnamefont{Schlabitz}},
  \bibinfo{author}{\bibfnamefont{A.}~\bibnamefont{Wurth}},
  \bibinfo{author}{\bibfnamefont{A.}~\bibnamefont{Mewis}}, \bibnamefont{and}
  \bibinfo{author}{\bibfnamefont{M.}~\bibnamefont{Reehuis}},
  \bibinfo{journal}{Physica B: Condensed Matter}
  \textbf{\bibinfo{volume}{252}}, \bibinfo{pages}{44} (\bibinfo{year}{1998}),
  ISSN \bibinfo{issn}{0921-4526},
  \urlprefix\url{https://www.sciencedirect.com/science/article/pii/S0921452697009046}.

\bibitem[{\citenamefont{Reehuis et~al.}(1998)\citenamefont{Reehuis, Jeitschko,
  Kotzyba, Zimmer, and Hu}}]{Reehuis1998}
\bibinfo{author}{\bibfnamefont{M.}~\bibnamefont{Reehuis}},
  \bibinfo{author}{\bibfnamefont{W.}~\bibnamefont{Jeitschko}},
  \bibinfo{author}{\bibfnamefont{G.}~\bibnamefont{Kotzyba}},
  \bibinfo{author}{\bibfnamefont{B.}~\bibnamefont{Zimmer}}, \bibnamefont{and}
  \bibinfo{author}{\bibfnamefont{X.}~\bibnamefont{Hu}},
  \bibinfo{journal}{Journal of Alloys and Compounds}
  \textbf{\bibinfo{volume}{266}}, \bibinfo{pages}{54} (\bibinfo{year}{1998}),
  ISSN \bibinfo{issn}{0925-8388},
  \urlprefix\url{https://www.sciencedirect.com/science/article/pii/S0925838897004866}.

\bibitem[{\citenamefont{Li and Hoffmann}(1986)}]{Li1986}
\bibinfo{author}{\bibfnamefont{J.}~\bibnamefont{Li}} \bibnamefont{and}
  \bibinfo{author}{\bibfnamefont{R.}~\bibnamefont{Hoffmann}},
  \bibinfo{journal}{Zeitschrift für Naturforschung B}
  \textbf{\bibinfo{volume}{41}}, \bibinfo{pages}{1399} (\bibinfo{year}{1986}),
  \urlprefix\url{https://doi.org/10.1515/znb-1986-1114}.

\bibitem[{Sup(2023)}]{Supplement}
\bibinfo{journal}{See supplement material for the relation among pressure
  dependent unit cell volume, $T{_N}$ and Eu valence, F-f plot of EuCoGe$_3$,
  and the equation of state fitting of neon}  (\bibinfo{year}{2023}).

\bibitem[{\citenamefont{Uchima et~al.}(2014{\natexlab{b}})\citenamefont{Uchima,
  Takaesu, Akamine, Kakihana, Tomori, Uejo, Teruya, Nakamura, Hedo, Nakama
  et~al.}}]{Uchima2014b}
\bibinfo{author}{\bibfnamefont{K.}~\bibnamefont{Uchima}},
  \bibinfo{author}{\bibfnamefont{Y.}~\bibnamefont{Takaesu}},
  \bibinfo{author}{\bibfnamefont{H.}~\bibnamefont{Akamine}},
  \bibinfo{author}{\bibfnamefont{M.}~\bibnamefont{Kakihana}},
  \bibinfo{author}{\bibfnamefont{K.}~\bibnamefont{Tomori}},
  \bibinfo{author}{\bibfnamefont{T.}~\bibnamefont{Uejo}},
  \bibinfo{author}{\bibfnamefont{A.}~\bibnamefont{Teruya}},
  \bibinfo{author}{\bibfnamefont{A.}~\bibnamefont{Nakamura}},
  \bibinfo{author}{\bibfnamefont{M.}~\bibnamefont{Hedo}},
  \bibinfo{author}{\bibfnamefont{T.}~\bibnamefont{Nakama}},
  \bibnamefont{et~al.}, \bibinfo{journal}{Journal of Physics: Conference
  Series} \textbf{\bibinfo{volume}{568}}, \bibinfo{pages}{042032}
  (\bibinfo{year}{2014}{\natexlab{b}}),
  \urlprefix\url{https://doi.org/10.1088/1742-6596/568/4/042032}.

\bibitem[{\citenamefont{Angel et~al.}(2014)\citenamefont{Angel,
  Gonzalez-Platas, and Alvaro}}]{EosFit}
\bibinfo{author}{\bibfnamefont{R.}~\bibnamefont{Angel}},
  \bibinfo{author}{\bibfnamefont{J.}~\bibnamefont{Gonzalez-Platas}},
  \bibnamefont{and} \bibinfo{author}{\bibfnamefont{M.}~\bibnamefont{Alvaro}},
  \bibinfo{journal}{Zeitschrift für Kristallographie}  (\bibinfo{year}{2014}).

\bibitem[{\citenamefont{Birch}(1947)}]{Birch1947}
\bibinfo{author}{\bibfnamefont{F.}~\bibnamefont{Birch}},
  \bibinfo{journal}{Phys. Rev.} \textbf{\bibinfo{volume}{71}},
  \bibinfo{pages}{809} (\bibinfo{year}{1947}),
  \urlprefix\url{https://link.aps.org/doi/10.1103/PhysRev.71.809}.

\bibitem[{\citenamefont{Heinz and Jeanloz}(1984)}]{EosGold_1984}
\bibinfo{author}{\bibfnamefont{D.~L.} \bibnamefont{Heinz}} \bibnamefont{and}
  \bibinfo{author}{\bibfnamefont{R.}~\bibnamefont{Jeanloz}},
  \bibinfo{journal}{Journal of Applied Physics} \textbf{\bibinfo{volume}{55}},
  \bibinfo{pages}{885} (\bibinfo{year}{1984}),
  \eprint{https://doi.org/10.1063/1.333139},
  \urlprefix\url{https://doi.org/10.1063/1.333139}.

\bibitem[{\citenamefont{Hemley et~al.}(1989)\citenamefont{Hemley, Zha,
  Jephcoat, Mao, Finger, and Cox}}]{Hemley_PRB_1989}
\bibinfo{author}{\bibfnamefont{R.~J.} \bibnamefont{Hemley}},
  \bibinfo{author}{\bibfnamefont{C.~S.} \bibnamefont{Zha}},
  \bibinfo{author}{\bibfnamefont{A.~P.} \bibnamefont{Jephcoat}},
  \bibinfo{author}{\bibfnamefont{H.~K.} \bibnamefont{Mao}},
  \bibinfo{author}{\bibfnamefont{L.~W.} \bibnamefont{Finger}},
  \bibnamefont{and} \bibinfo{author}{\bibfnamefont{D.~E.} \bibnamefont{Cox}},
  \bibinfo{journal}{Phys. Rev. B} \textbf{\bibinfo{volume}{39}},
  \bibinfo{pages}{11820} (\bibinfo{year}{1989}),
  \urlprefix\url{https://link.aps.org/doi/10.1103/PhysRevB.39.11820}.

\bibitem[{\citenamefont{Momma and Izumi}(2011)}]{Momma2011}
\bibinfo{author}{\bibfnamefont{K.}~\bibnamefont{Momma}} \bibnamefont{and}
  \bibinfo{author}{\bibfnamefont{F.}~\bibnamefont{Izumi}},
  \bibinfo{journal}{Journal of Applied Crystallography}
  \textbf{\bibinfo{volume}{44}}, \bibinfo{pages}{1272} (\bibinfo{year}{2011}),
  \urlprefix\url{https://doi.org/10.1107/S0021889811038970}.

\end{thebibliography}

\end{document}